\documentclass{aa}

\usepackage{graphicx}

\usepackage{txfonts}
\usepackage{threeparttablex}
\usepackage{supertabular,booktabs}

\usepackage{chemformula} 
\usepackage{longtable}
\titlerunning{Chemical modelling of interstellar MgS} 
\authorrunning{Rey-Montejo et al.} 

\title{Chemical modelling of interstellar MgS}
\author{
    Marta Rey-Montejo\inst{1,2}, 
    Izaskun Jim{\'e}nez-Serra\inst{1}, Tom Millar\inst{3}, Ryan C. Fortenberry\inst{4}, Serena Viti\inst{5,6,7}, Juan Garc{\'i}a de la Concepci{\'o}n\inst{8}, Gijs Vermariën\inst{5,9}, Miguel Sanz-Novo\inst{10}, Laura Colzi\inst{1}, Shaoshan Zeng\inst{11}, V{\'i}ctor M. Rivilla\inst{1}
}

\institute{
    Centro de Astrobiología (CAB), CSIC-INTA, Ctra. de Ajalvir km 4, E-28850, Torrejón de Ardoz, Spain \\
    \email{mrey@cab.inta-csic.es}
    \and
    Departamento de F{\'i}sica de la Tierra y Astrof{\'i}sica, Facultad de Ciencias F{\'i}sicas, Universidad Complutense de Madrid, 28040 Madrid, Spain
    \and
    Astrophysics Research Centre, School of Mathematics and Physics, Queen’s University Belfast, University Road, Belfast BT71NN, UK
    \and
    Department of Chemistry \& Biochemistry, University of Mississippi, University, MS 38677-1848, USA
    \and
    Leiden Observatory, Leiden University, PO Box 9513, NL-2300 RA Leiden, the Netherlands
    \and 
    Transdisciplinary Research Area (TRA) 'Matter'/Argelander-Institut f\"ur Astronomie, University of Bonn, Germany
    \and
    Physics and Astronomy, University College London, Gower Street, London WC1E 6BT, UK 
    \and
    Departamento de Química Orgánica e Inorgánica, Facultad de Ciencias, and IACYS-Green Chemistry and Sustainable Development Unit, Universidad de Extremadura, Badajoz, Spain
    \and
    SURF, Amsterdam, The Netherlands
    \and
    Center for Astrochemical Studies, Max-Planck-Institut f\"ur extraterrestrische Physik, Giessenbachstrasse 1, Garching bei Munchen, 85748, Germany
    \and
    Star and Planet Formation Laboratory, RIKEN Pioneering Research Institute, RIKEN, 2–1 Hirosawa, Wako, Saitama, 351–0198, Japan
}

\date{Received May 28, 2026; accepted July 27, 2026}
\begin{document}

  \abstract
   {The detection of magnesium sulphide (MgS) and sodium sulphide (NaS) towards the Galactic Center molecular cloud G+0.693 constitutes the first detection of metal sulphides in the interstellar medium (ISM). However, there is scarce information about the key reactions (either in the gas phase or on grains) involved in their formation.} 
   {In this paper, we model the chemistry of MgS simulating the passage of a low-velocity shock to recover the abundances recently measured towards G+0.693. Through this chemical modelling, we analyse the dominant reactions involved in the formation and destruction of this molecule, their associated chemical timescales, and the depletion factor needed to recover the observed abundances.}
   {We build the initial chemical network of MgS by using SiS as a proxy for this metal sulphide, and we investigate the exothermicity of these and additional, uniquely proposed reactions through quantum chemical computations. We run a three-phase model (initial translucent cloud, cloud collapse phase, and shock interaction stage) that mimics the evolution and physical conditions of G+0.693.}
   {Our results show that a depletion factor of 1000 is required for elemental Mg to recover the observed abundances of MgS. This implies that potentially more than 99.9\% of Mg is locked in dust grains. The dominant reaction leading to the formation of MgS is the neutral-neutral reaction between MgH and S in the gas phase.}
   {This work represents the first analysis of the chemistry of the metal-sulphide MgS and suggests that Mg is largely incorporated into dust grains, most likely in the form of silicates. However, additional laboratory and/or theoretical studies of the key MgS formation reactions are essential to obtain more reliable constraints. Future missions, such as PRIMA, will provide  insights into the number of metal sulphides locked into interstellar dust grains.}


   \maketitle
%

\section{Introduction}
The Galactic Center molecular cloud G+0.693-0.027 (hereafter G+0.693) is distinguished by its chemically-rich environment, where a wide range of molecules, including complex organic molecules (COMs)\footnote{Complex organic molecules are generally referred to carbon containing molecules with six or more atoms.}, have been detected, some of them for the first time \citep[e.g.][]{Izaskun2020, Izaskun2022, Rivilla2019, Rivilla2020, Rivilla2021a, Rivilla2021b, Rivilla2022a, Rivilla2022b, Rivilla2023, Rivilla2026, Rodriguez2021b, Miguel2023, Miguel2024a, Miguel2024b, Miguel2025, Zeng2021, Zeng2023, SAndres2024, RM2024, Araki2026}. This enhanced chemical complexity has been attributed to a cloud-cloud collision, which induces low-velocity shocks that release significant amounts of material from dust grains into the gas phase through sputtering \citep{Zeng2020}.\\

Recent work has shown the detection of sodium sulphide (NaS) and magnesium sulphide (MgS) in the interstellar medium (ISM) towards G+0.693 \citep{RM2024}. These diatomic metal-sulphide molecules suggest a connection between sulphur (S) and metals, indicating that S may be depleted onto dust grains since metals are known to be heavily depleted in dense molecular clouds \citep{Field1974, Savage1996, Savaglio2003, Jenkins2009, DCia2015, Roman2021, Konstant2023, Konstant2024}. This may shed light on the so-called sulphur depletion problem: the abundances of S-bearing volatile species such as H$_2$S, SO, and SO$_2$ ($\sim$10$^{-10}$, with respect to molecular H$_2$) in molecular clouds are orders of magnitude lower than the atomic S cosmic abundance \citep[1.32$\times$10$^{-5}$;][]{Asplund2009}. 

The intrinsic nature of these metal-bearing molecules makes their detection particularly challenging, as they are difficult to maintain in the gas phase in the ISM.
This has, therefore, hindered the study of their chemistry in the ISM. On top of that, there are scarce laboratory experiments to reproduce the chemical reactions needed to form these molecules under ISM conditions.\\
 
In this work, we focus on MgS only, because some theoretical work has been carried out already for this molecule \citep{wei25,bell25}. 
Previous studies indicate that atomic magnesium (Mg) remains ionised even at large optical depths, suggesting that MgS dust can easily form through direct nucleation from the gas phase in evolved stars \citep{Kimura2005}. The far-IR band of solid MgS at $\sim$30 $\mu$m \citep[which is highly sensitive to the shape and/or surface properties of the MgS grains; see ][]{Kimura2005} has been proposed as the main carrier of the 30 $\mu$m feature observed towards carbon-rich AGB (Asymptotic Giant Branch) stars \citep{Goebel1985}. As such, solid MgS could become part of the refractory core of circumstellar dust, which is incorporated afterwards into the ISM. Alternatively, diatomic MgS may form via gas-phase reactions in molecular clouds following reaction mechanisms similar to those of other Mg-bearing molecules such as metal cyanides and acetylides. These mechanisms start from the radiative association of Mg$^+$ with large cyanopolyynsues and polyynes, and undergo the subsequent dissociative recombination of the corresponding Mg$^+$/NC$_{2n+1}$H and Mg$^+$/C$_{2n}$H$_2$ complexes with electrons \citep{Cernicharo2023}. Recent investigations have also shown that the radiative association of Mg + S \citep{wei25}, and the neutral-neutral reaction of MgH with HS \citep{bell25} yield MgS.\\

Given the limited information currently available, the chemistry of these metal-bearing molecules must be analysed to understand how they form in the ISM. Such an analysis will help to better constrain their molecular abundances and the composition of dust grains, which play a crucial role in the star formation process and in the final chemical composition of planetesimals and cometesimals.\\ 

In this paper, we used the chemical code UCLCHEM \citep{Holdship2017} to model the chemistry of MgS simulating the physical conditions and the low-velocity shock affecting the Galactic Center molecular cloud G+0.693. 
The article is organised as follows. Section~\ref{sec:chemical_modelling} describes the chemical network of MgS and the physical conditions that reproduce the environment in G+0.693. In Section~\ref{sec:results}, we show the results of the chemical model including the depletion factors derived for S and Mg and the key reaction(s) leading to the formation of MgS. In Section~\ref{sec:discussion} we analyse the stage at which the reaction dominates the production of MgS and the implication of these results, and discuss the main findings of our study. Finally, Section~\ref{sec:conclusion} presents our conclusions.

\section{Chemical modelling}\label{sec:chemical_modelling}
\subsection{Chemical network}
Silicates are the predominant components of the refractory cores in interstellar dust \citep{Fabian2001, Draine2003}. 
The cosmic abundance ratio of magnesium and silicon (Si) is Mg/Si $\approx$ 1 \citep{Turner1985}. Additionally, both elements are expected to be depleted on grains, so SiS - much more studied - is used as an initial proxy for reactions involving Mg, as it provides a starting point in the absence of available information.
Of course, the chemistries of these elements differ, and the unique chemistry of Mg is explored in this work.\\

For silicon-bearing species, SiO is either released directly from grains in shocks \citep{Martin1992, Izaskun2008} or formed in the gas phase following the liberation of atomic Si \citep{Schilke1997, Gusdorf2008}. Differently, the formation of SiS has been studied primarily in the gas phase. After the release of atomic Si in a shock, SiS can form through multiple pathways: i) ion-neutral gas-phase processes involving Si$^+$; ii) neutral-neutral gas-phase reactions with atomic Si or molecules such as SiH$_n$ (with n = 1, 2, 3, 4) \citep[e.g. Si + S/S$_2$/HS and SiH$_n$ + S/S$_2$/HS][]{Podio2017, Rosi2018, Fortenberry2024}; iii) through the reaction S$^+$($^4$S) + SiH$_2$($^1$A$_1$) \cite[see][]{Mancini2022}; or iv) through the interaction of atomic Si with either SO or SO$_2$ \citep{Zanchet2018, Campanha2022}. \\

For the MgS chemical network, we used the default dust-grain network of UCLCHEM, based on \citet{Quenard2018}, and the UMIST22 database \citep{Millar2024} for the gas-phase network. 
We first duplicated the chemical network of SiS and modified it to reproduce the formation of MgS by replacing Si with Mg in each reaction. The reactions in our network obtained by duplicating the SiS network can be found in Tables~\ref{tab:reac_MGS} and~\ref{tab:chem_mgs_grain}. At a first stage, we inspected the exothermicity of these reactions by using the NIST\footnote{\url{https://webbook.nist.gov/chemistry/}} database. We then performed a few initial modelling runs with UCLCHEM to identify the dominant formation reactions of MgS, which were subsequently explored in detail using quantum chemical computations to obtain their potential energy surfaces (PES) (Section~\ref{sec:quantum_calculations}). Several reactions such as Mg + H$_2$S $\rightarrow$ MgS + H$_2$ or Mg$^+$ + OCS $\rightarrow$ MgS$^+$ + CO were found to be largely endothermic (Table~\ref{tab:second_stage_reac_MGS}) and therefore their $\alpha$, $\beta$ and $\gamma$ parameters\footnote{These parameters follow the Kooij-Arrhenius equation $k(\rm T) = \alpha (\frac{T}{300K})^{\beta}\rm exp (-\gamma/T)$, where T corresponds to the temperature. $\alpha$ is the pre‑exponential factor which sets the overall scale of the rate constant. $\beta$ is the temperature‑dependence exponent that controls how strongly the rate changes with temperature. $\gamma$ is the activation energy term, and represents the energy barrier.} were set to zero (see Table~\ref{tab:reac_MGS}). A list of new reactions were additionally explored, as reported in Table~\ref{tab:new_reac_MGS}. While the majority of them were also found endothermic (Table~\ref{tab:new_reac_MGS}), some exceptions arose, including Mg + SH$^+$ $\rightarrow$ MgS$^+$ + H and MgH + S $\rightarrow$ MgS + H.   
 
Recent work has studied additional reactions for the formation of MgS in the gas phase. \cite{bell25} carried out quantum chemical computations that show that the neutral-neutral reaction of MgH + HS is an energetically downhill pathway leading to the MgS molecule as well as H$_2$. By using the information provided in \cite{bell25}, we have calculated the global rate constant for this reaction (see Section \ref{sec:constan_rates}) and added this reaction to the network (Table~\ref{tab:reac_MGS}).
Moreover, \cite{wei25} proposed the formation of MgS through radiative association occurring during collisions between Mg and S atoms. They employed quantum mechanical ab initio calculations to predict the association rate coefficient for the formation of MgS. Their computational results show that, within the temperature range of 10–10000 K, the 1$^1\Pi$ → X$^1\sum^+$ transition dominates the formation of MgS via radiative association in collisions between Mg($^1$S) and S($^1$D) atoms, while the 1$^1\Lambda$ → 1$^1\Pi$ transition becomes significant at higher temperatures. In our study, the maximum shock temperature does not exceed 1000 K; therefore, we employed the $\alpha$, $\beta$ and $\gamma$ parameters derived for the 1$^1\Pi$ → X$^1\sum^+$ transition (see Table~\ref{tab:reac_MGS}). The final gas-phase chemical network includes 494 species and 8885 reactions.\\

\subsection{Physical conditions}
We used the UCLCHEM\footnote{\url{https://uclchem.github.io/v3.5.5/}} (v3.5.3) astrochemical code \citep{Holdship2017} to simulate the physical conditions and the low-velocity shock interaction present in the G+0.693 cloud in the Galactic Center.
The model has been run in three different phases to reproduce the physical conditions in G+0.693, as it was previously carried out by \cite{Rivilla2022b} and \cite{Miguel2024a}: Phase 0 simulates the chemistry of a translucent cloud\footnote{The cloud model reproduces spherical clouds of gas with constant physical conditions. The exception to this is the density which can be modified using the freefall parameter.} with a final visual extinction\footnote{The final visual extinction is calculated as: \[\text{A$_{\rm v}$} = A_{\rm v_0} + \frac{n(\rm H)\times R_{\rm out}}{1.6\times10^{21}} [\rm mag],\] with R$_{\rm out}$ in cm.} of 1.9 mag\footnote{To reproduce a translucent cloud we used a initial visual extinction ($A_{\rm v_0}$) of 1 mag and R$_{\rm out}$=0.5 pc.}, exhibiting constant $n(\rm H)$=10$^3$ cm$^{-3}$ and $T_{\rm kin}$=20 K for a period of 10$^6$ years. The initial elemental abundances used in the model are listed in Table~\ref{tab:elemental_abund}. These values represent the volatile fraction of each element, i.e. the portion available to participate in gas–ice chemistry. UCLCHEM uses these elemental abundances as the starting point for phase 0, and the chemical network evolves them into molecular abundances through gas-phase and grain-surface processes. At the end of each phase, the resulting molecular abundances are carried forward as the initial conditions for the subsequent phase, ensuring chemical continuity throughout the physical evolution. For models for which the initial abundance of one element is depleted, the depleted fraction is locked onto grain cores and it does not participate further of the subsequent gas-ice chemistry. This means that the remaining volatile abundance is depleted from the start of phase 0, so the chemical evolution reflects only the fraction of the element available to interact between the gas and ice phases.

\begin{table}[h]
    \centering
    \begin{ThreePartTable}
    \caption{Elemental abundances used by UCLCHEM.}
    \begin{tabular}{c c}
    \toprule
    Atom & Abundance\\
    \midrule
    \midrule
        O & 3.34$\times10^{-4}$\\
        N & 6.18$\times10^{-5}$\\
        S & 1.32$\times10^{-5}$\\
        Mg & 3.98$\times10^{-5}$\\
        Si & 1.78$\times10^{-6}$\\
        \bottomrule
    \end{tabular}
    \label{tab:elemental_abund}
        \begin{tablenotes}
           \item Note. Elemental abundances are taken from \cite{Asplund2009}.
        \end{tablenotes}
    \end{ThreePartTable}
\end{table}

In Phase 1 we simulate the freefall collapse of a molecular cloud using the freefall parameter to reach the initial density of the shock phase (Phase 2). The final abundances of Phase 1 are then used as the starting abundances in the shock model. In this way, in Phase 1 the cloud evolves from a initial density of $n(\rm H)$=10$^3$ cm$^{-3}$ to a final density of $n(\rm H)$=2$\times$10$^4$ cm$^{-3}$, with a final visual extinction of 29 mag\footnote{To reproduce a dense cloud we used $A_{\rm v_0}$=2 mag and R$_{\rm out}$=1 pc.}, while $T_{\rm kin}$ is set constant to 10 K . 

\begin{figure}[h]
    \centering
    \includegraphics[width=\linewidth]{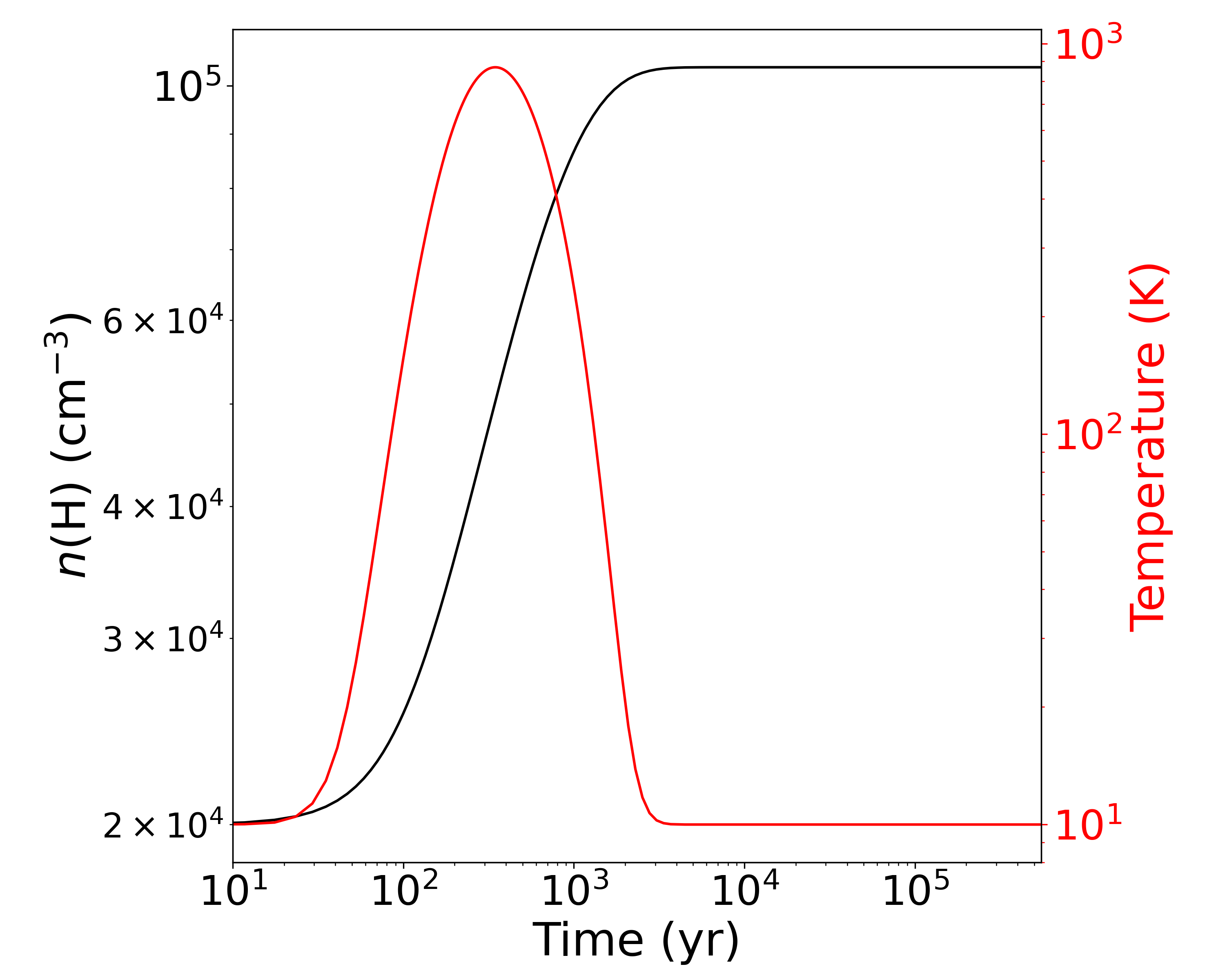}
    \caption{Physical conditions of phase 2: We consider for this phase a C-type shock with a shock speed of $c_s$=20 km s$^{-1}$ and an initial H volume gas density of $n(\rm H)$=2$\times$10$^4$ cm$^{-3}$, where $n(\rm H)$=2$\times n(\rm H_2)$. The red line indicates the evolution of the temperature of the neutral fluid within the shock. The black line shows the evolution of the H volume density.}
    \label{fig:density}
\end{figure}

Finally, Phase 2 simulates the passage of a low-velocity C-type shock to calculate the time dependent evolution of the molecular abundances sputtered from grains. The C-shock model parametrises the density, temperature and velocity profiles of C-shocks as a function of shock velocity, initial gas density and magnetic field \citep{Izaskun2008,Holdship2017}. 
We adopt a shock velocity of $v_s$=20 km s$^{-1}$, consistent with the line widths of the molecular line emission observed in G+0.693 \citep{Requena2006, Zeng2018}, and a H volume gas density of $n(\rm H)$=2$\times$10$^4$ cm$^{-3}$ similar to that measured towards this cloud \citep{Zeng2020, colzi24}. During the shock, the gas temperature reaches values close to 1000 K and progressively cools down as the shock progresses and compresses the gas to the final H volume density of a few times 10$^5$ cm$^{-3}$. After the passage of the shock, the temperature returns to its initial value of 10 K (see Figure~\ref{fig:density}). The cooling of the gas is consistent with the enhanced D/H ratios measured for molecules such as HCO$^+$, N$_2$H$^+$, HCN and HNC towards G+0.693 \citep{colzi22}. The enhanced D/H ratios indicate the presence of a cooler, postshocked-gas core with T$_{\rm kin}$ $\leq$30 K \citep[i.e. lower than the typical value of 70–140 K;][]{Krieger2017, Zeng2018}, and a H$_2$ volume density of $n(H_2)\geq$5$\times$10$^4$ cm$^{-3}$ \citep{colzi24}, similar to the H volume gas density reached in the shock (see Figure~\ref{fig:density}).

\section{Result}\label{sec:results}

\subsection{Constraining the amount of sulphur depleted on dust grains in G+0.693}\label{sec:sulphur_constrain}
Reproducing the observed abundances of MgS is a highly-degenerate problem because of the various parameters involved in the astrochemical modelling.
First, G+0.693 is believed to be influenced by a strong secondary ultraviolet (UV) radiation field generated by the collision of cosmic-rays with H$_2$ molecules \citep{Rivilla2023}. As such, the cosmic-ray ionisation rates must be constrained. In addition, the depletion factor of S must also be established in order to be able to determine the depletion needed for Mg to reproduce the observed abundances of MgS. Finally, the typical timescales of the shocked gas in Galactic Center clouds are needed to compare the observed abundances with our modelling results.

To cover the different possibilities for all these parameters, we first ran the model with different depletion factors of S (1, 10 and 100) with respect to the elemental abundance reported by \cite{Asplund2009}, of 1.32$\times$10$^{-5}$. To constrain the S depletion factor, we compared the modelling results with the observed abundances of the S-bearing molecules SO, OCS, NS, SiS and HCS$^+$ towards G+0.693 (see Table~\ref{tab:abund_S_MG}). These molecules were selected because their abundances are well constrained in this source, providing a robust benchmark that the models must reproduce.
We did this for different cosmic-ray ionisation rates: the standard Galactic disk value of $\zeta$=1.3$\times$10$^{-17}$ s$^{-1}$, as well as two $\zeta$ enhanced values of 100 times and 1000 times higher than the standard one (see Figure~\ref{fig:S_constrain_todos}). The higher values have been invoked in the chemical modelling of ions such as HOCS$^+$ and PO$^+$ in G+0.693 \citep{Rivilla2022b,Miguel2024a}. 

\begin{figure}[h]
    \centering
    \includegraphics[width=\linewidth]{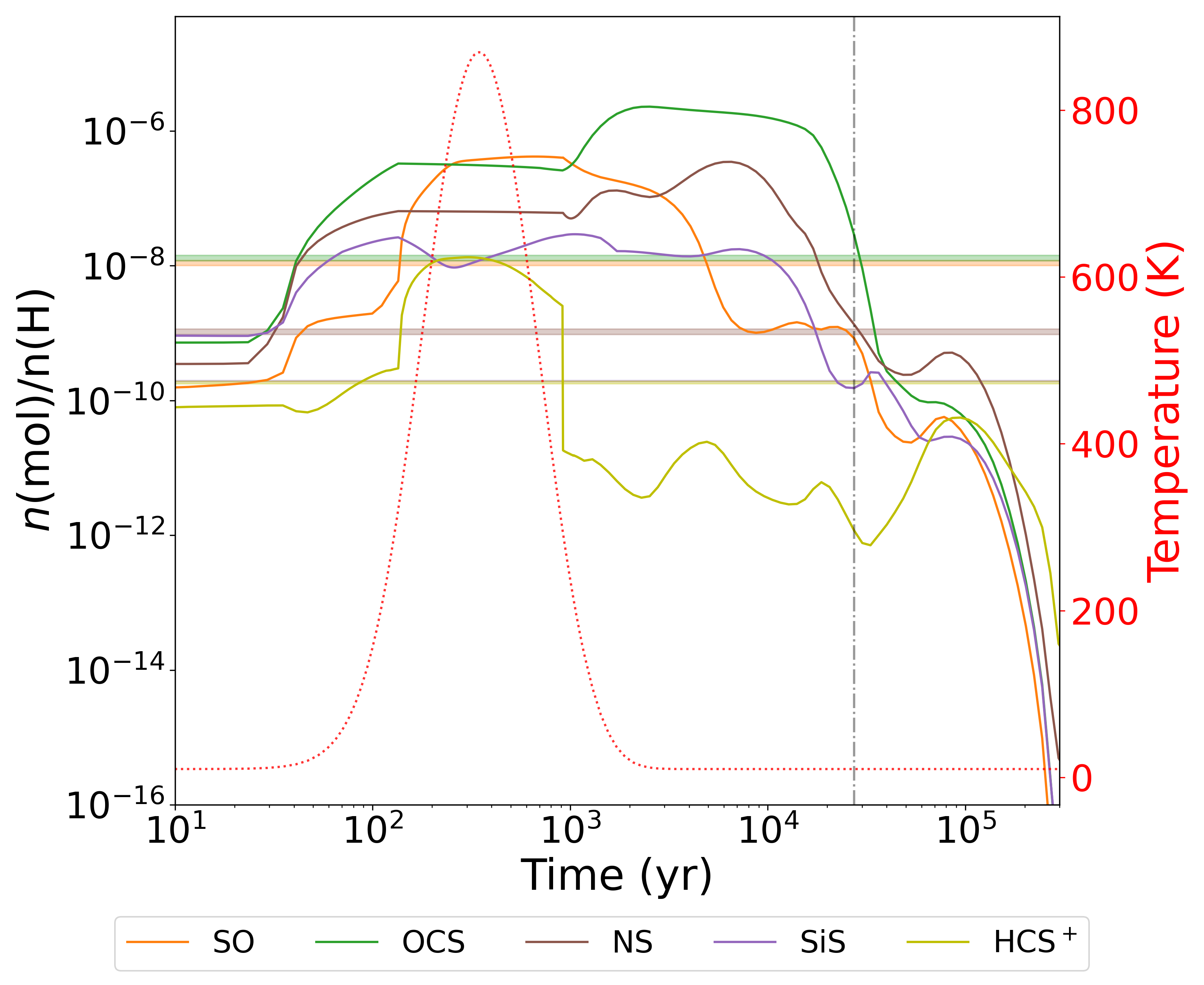}
    \caption{Chemical model results (Phase 2) for $\zeta$=1.3$\times$10$^{-15}$ and an S depletion factor of 1 (i.e, no depletion): evolution of the fractional abundances of SO, OCS, NS, SiS and HCS$^+$ in the gas phase as a function of time (colour lines). 
    The vertical grey dash-dotted line corresponds to $\sim$2.7$\times$10$^4$ years as inferred using the MAE method and indicates the timescale at which the minimum MAE value is reached. The red dashed curve indicates the evolution of the temperature of the neutral fluid in the shock. The colour-shaded horizontal lines correspond to the observed abundances (SO, OCS, NS, SiS and HCS$^+$) measured towards G+0.693, and converted into abundance with respect to n(H) by adopting $n$(H)=2$\times$$n$(H$_2$). The width of the horizontal shaded-lines consider the error in the measured abundances, and are colour-coded according to the legend at the bottom of the figure.}
    \label{fig:S_constrain}
\end{figure}

\begin{table*}[h]
    \centering
    \begin{ThreePartTable}
    \caption{Observed abundances of the sulphur-bearing molecules towards G+0.693.}
    \begin{tabular}{c c c c c c c}
        \toprule
        {Molecule} & ${N}$ & ${T_{\rm ex}}$ & ${v_{\rm LSR}}$ & {FWHM} & {Abundance$^a$} & {Ref.} \\
        & ($\times 10^{14}$ cm$^{-2}$) & (K) & (km s$^{-1}$) & (km s$^{-1}$) & ($\times 10^{-9}$) & \\
        \midrule
        \midrule
            SO & 30.06$\pm$0.07 & 6.9$^b$ & 67.9$\pm$0.3 & 24.8$\pm$0.7 & 11.0$\pm$0.01 & (1)\\
            SiS & 0.53$\pm$0.01 & 8.0$\pm$0.1 & 66.8$\pm$0.2 & 24.0$\pm$0.3 & 0.195$\pm$ 0.004 & (2)\\
            OCS & 36.1$\pm$0.5 & 22.9$\pm$0.3 & 66.8$\pm$0.1 & 21.8$\pm$0.3 & 13.0$\pm$1.2 & (3)\\
            NS & 2.8$\pm$0.3 & 7.5$\pm$0.5 & 70.6$\pm$0.6 & 19$\pm$1 & 1.05$\pm$0.1 & (4)\\
            HCS$^+$ & 0.53$\pm$0.2 &6.9$\pm$0.2 & 68.5$\pm$0.2 & 20.0$\pm$0.6 & 0.02$\pm$0.07 & (3)\\
            MgS & 0.00060$\pm$0.00006 & 39$\pm$5 & 69.0$^b$ & 15$^b$ & 0.00023$\pm$0.00003 & (5)\\
        \bottomrule
    \end{tabular}
    \begin{tablenotes}
    \footnotesize
    \item[a] We adopted $N_{\rm H}$=2$\times N_{\rm H_2}$, with $N_{\rm H_2}$=1.35$\times$10$^{23}$cm$^{-2}$ from \cite{martin2008}.
    \item[b]  Value fixed in the fit.
    \item Ref. (1) \cite{Rivilla2022b}; (2) \cite{Massalkhi2023}; (3) \cite{Miguel2024a}; (4) \cite{Miguel2024b}; (5) \cite{RM2024}.
    \end{tablenotes}
    \label{tab:abund_S_MG}
    \end{ThreePartTable}
\end{table*}
To establish the best-fitting model, the mean absolute error (MAE\footnote{The MAE is calculated as: \[\text{MAE} = \frac{1}{N} \sum_{i=1}^{N} \left| \log_{10}(\hat{y}_i) - \log_{10}(y_i) \right|,\] where $y_i$ represents the observed value of species i (in abundance with respect to H), $\hat{y}_i$ is the value predicted by the astrochemical model, and N is the number of species being compared.}) was calculated between the predicted and observed abundances of all the above S-species on a logarithmic scale following \citet{Wakelam2006} and \citet{Izaskun2025}. In this calculation, we also took into account the associated uncertainties derived in Sections~\ref{sec:s_analy} and~\ref{sec:mg_analy} using the Python library RichValues\footnote{\url{https://pypi.org/project/richvalues  }} \citep{Andres2025}, and that arise from the error propagation of the observational data in the MAE analysis.
We imposed that the MAE is calculated at timescales beyond the end of the passage of the shock so that other minima reached at the beginning of the simulation are discarded.  From this analysis we found that the lowest MAE is obtained for a depletion factor of 1 (i.e. no depletion of S) and a cosmic-ray ionisation rate of $\zeta$=1.3$\times$10$^{-15}$ s$^{-1}$ (enhanced by a factor of 100; see Section~\ref{sec:s_analy}),  at a timescale of the post-shock phase of $\sim$2.7$\times$10$^4$ years (see Figure~\ref{fig:S_constrain}) .

\subsection{Depletion factor for magnesium initial abundance}\label{sec:MgS}
After fixing the S depletion factor to 1 and the cosmic-ray ionisation rate to $\zeta$=1.3$\times$10$^{-15}$ s$^{-1}$, the model was run varying the depletion factor of Mg by factors of 1, 10, 100, 1000 and 10000 (see Figure~\ref{fig:Mg_constrain}) with respect to the initial elemental abundance reported for Mg by \cite{Asplund2009} of 3.98$\times$10$^{-5}$.

We then applied the MAE analysis to determine the best fitting model, taking into account the abundances of SO, OCS, NS, SiS, HCS$^+$ and MgS measured in G+0.693 (see Table~\ref{tab:abund_S_MG}), to ensure that the modelled abundances were consistent with the observed ones. 

\begin{figure}[h]
    \centering
    \includegraphics[width=\linewidth]{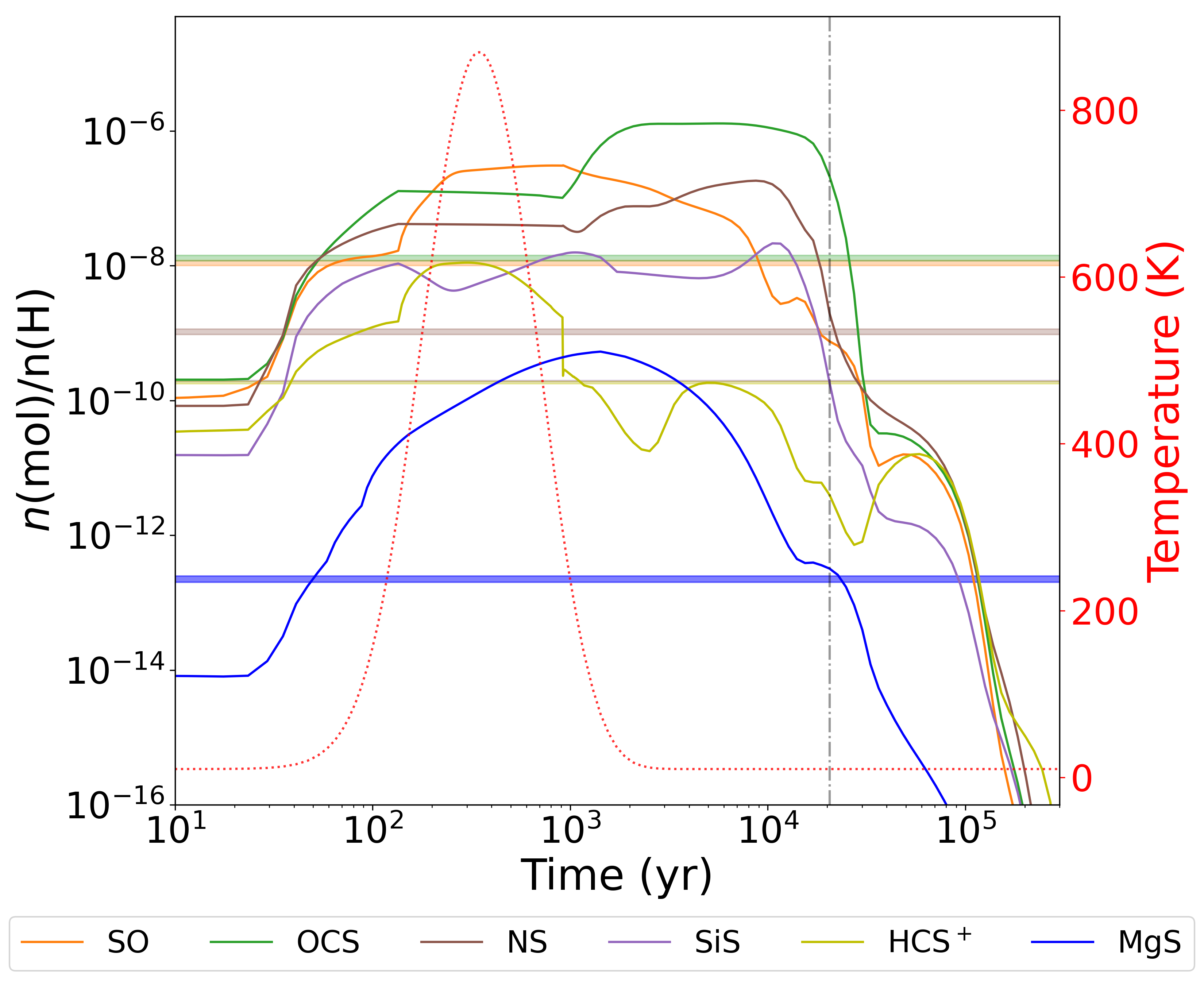}
    \caption{Chemical model results (Phase 2) for $\zeta$=1.3$\times$10$^{-15}$, no S depletion and Mg depletion factor of 1000: evolution of the fractional abundances of SO, OCS, NS, SiS, HCS$^+$ and MgS in the gas phase as a function of time.  
    The grey dash-dotted line corresponds to $\sim$2.1$\times$10$^4$ years and indicates the timescale when the minimum MAE value has been obtained after comparing with the observed molecular abundances. Finally, the colour-shaded horizontal lines correspond to the observed molecular abundances measured towards G+0.693, taking into account their error, and are colour-coded according to the legend at the bottom of the figure.}
    \label{fig:Deplet_Mg}
\end{figure}

The model with a Mg depletion factor of 1000 was the one that best fits the observed abundances (see Section~\ref{sec:mg_analy}), at a timescale of  $\sim$2.1$\times$10$^4$ years (see Figure~\ref{fig:Deplet_Mg}).
Note that the timescale derived is slightly shorter because we have added MgS to the other S species that were considered in the previous analysis (Section~\ref{sec:sulphur_constrain}).

\subsection{Main reactions driving the formation and destruction of MgS}\label{subsec:chem}

In UCLCHEM, the abundance of each species evolves according to differential equations that sum the contributions from all reactions that form or destroy it. Each reaction adds a positive term if it produces the species, or a negative term if it removes it, scaled by the reaction rate and the reactant abundances. By examining these individual contributions, the reactions that dominate the behaviour of MgS under the physical conditions of our model can be identified for the derived depletion factors of S and Mg. This will allow us to inspect the dominant chemical reactions involved in the production and destruction of MgS.

Table~\ref{tab:MgS} shows the ten most significant reactions involved in the production and destruction of MgS along the passage of the shock (Phase 2), for the best fitting model, this is, undepleted S (1.32$\times$10$^{-5}$), cosmic-ray ionisation rate of $\zeta$=1.3$\times$10$^{-15}$ s$^{-1}$ and magnesium abundance depleted by a factor of 1000 (3.98$\times$10$^{-8}$). The table shows the percentage contribution of each reaction to the net production or destruction of MgS.

\begin{table}[h]
    \centering
    \caption{Chemical reactions and their percentage of importance over the shock phase, divided into formation and destruction of MgS.}
    \begin{tabular}{l c}
        \toprule
        {Reaction} & {Percentage \%} \\
        \midrule
        \multicolumn{2}{c}{{Formation Reactions}} \\
        \midrule
        \ch{MgH + S -> MgS + H} & 90.2 \\
        \ch{MgSH$_2$ + C -> MgS + CH$_2$} & 5.6 \\
        \ch{MgH + HS-> MgS + H$_2$} & 1.6 \\
        \ch{HMgS$^+$ + NH$_3$ -> MgS + NH$_4 ^+$} & 0.87\\
        \ch{MgH + HCCS -> MgS + CCH$_2$} & 0.67 \\
        \ch{MgH + HCCS -> MgS + HCCH} & 0.67 \\
        \ch{HMgS$^+$ + HCN -> MgS + HCNH$^+$} & 0.24 \\
        \ch{MgS$^+$ + e$^-$ -> MgS + $hv$} & 0.04 \\
        \ch{MgH + S$_2$ -> MgS + HS} & 0.04 \\
        $^a$\ch{\#MgS + THERM -> MgS} & 0.02\\
        \midrule
        \multicolumn{2}{c}{{Destruction Reactions}} \\
        \midrule
        \ch{H$_3$$^+$ + MgS -> HMgS+ + H$_2$} & 46.5\\
        \ch{S$^+$ + MgS -> MgS$^+$ + S} & 13.5 \\
        \ch{H$^+$ + MgS -> MgS$^+$ + H} & 9.9 \\
        \ch{HCO+ + MgS -> HMgS+ + CO} & 7.5 \\
        \ch{He$^+$ + MgS -> S$^+$ + Mg + He} & 7.1 \\
        \ch{He$^+$ + MgS -> S + Mg$^+$ + He} & 7.1 \\
        \ch{MgS + CRPHOT -> S + Mg} & 4.4 \\
        \ch{C$^+$ + MgS -> MgS$^+$ + S} & 1.6 \\
        \ch{C$^+$ + MgS -> MgC$^+$ + C} & 1.6 \\
        \ch{MgS + FREEZE -> \#MgS} & 0.68 \\
        \bottomrule
    \end{tabular}
    \label{tab:MgS}
\begin{tablenotes}
\footnotesize
\item $^a$ \#MgS denotes surface ice MgS.
\end{tablenotes}
\end{table}

\section{Discussion}\label{sec:discussion}
\subsection{Constraining the sulphur depletion factor in G+0.693}
Determining the abundance of S is particularly difficult as it cannot be observed directly in the radio. As a result, estimates of the elemental S abundance in molecular clouds rely on observations of less abundant S-bearing species, such as SO, OCS or CS, along with the predictive capabilities of astrochemical models. \\

The S abundance towards other astronomical sources is uncertain and it likely depends upon the environment \citep{Fuente2023}. While the abundances of S-bearing species are consistent with undepleted S in diffuse clouds \citep{Neufeld2015} and highly irradiated photon-dominated regions such as the high-mass star-forming cluster of Orion A \citep{Fuente2023}, a depletion factor of $\sim$20 is required to explain those S abundances observed from low-mass to high-mass star-forming regions, such as Taurus and Perseus, respectively \citep{Fuente2023, Alvaro2024}.\\

The results of Section~\ref{sec:sulphur_constrain} above show that no S depletion is needed to reproduce the abundances of the S-bearing molecules SO, OCS, NS, SiS and HCS$^+$ towards G+0.693. This sets the initial abundance of S in our models to 1.32$\times$10$^{-5}$ \citep{Asplund2009}. This result is consistent with previous works \citep[][]{Rivilla2022b, Miguel2024a} and falls within the range of 0.7–3.5 $\times$ 10$^{-5}$ as reported for several sources in the Galactic Center \citep{RodriguezMartin2005}. 

\begin{figure*}[h]
    \centering
    \includegraphics[width=0.9\linewidth]{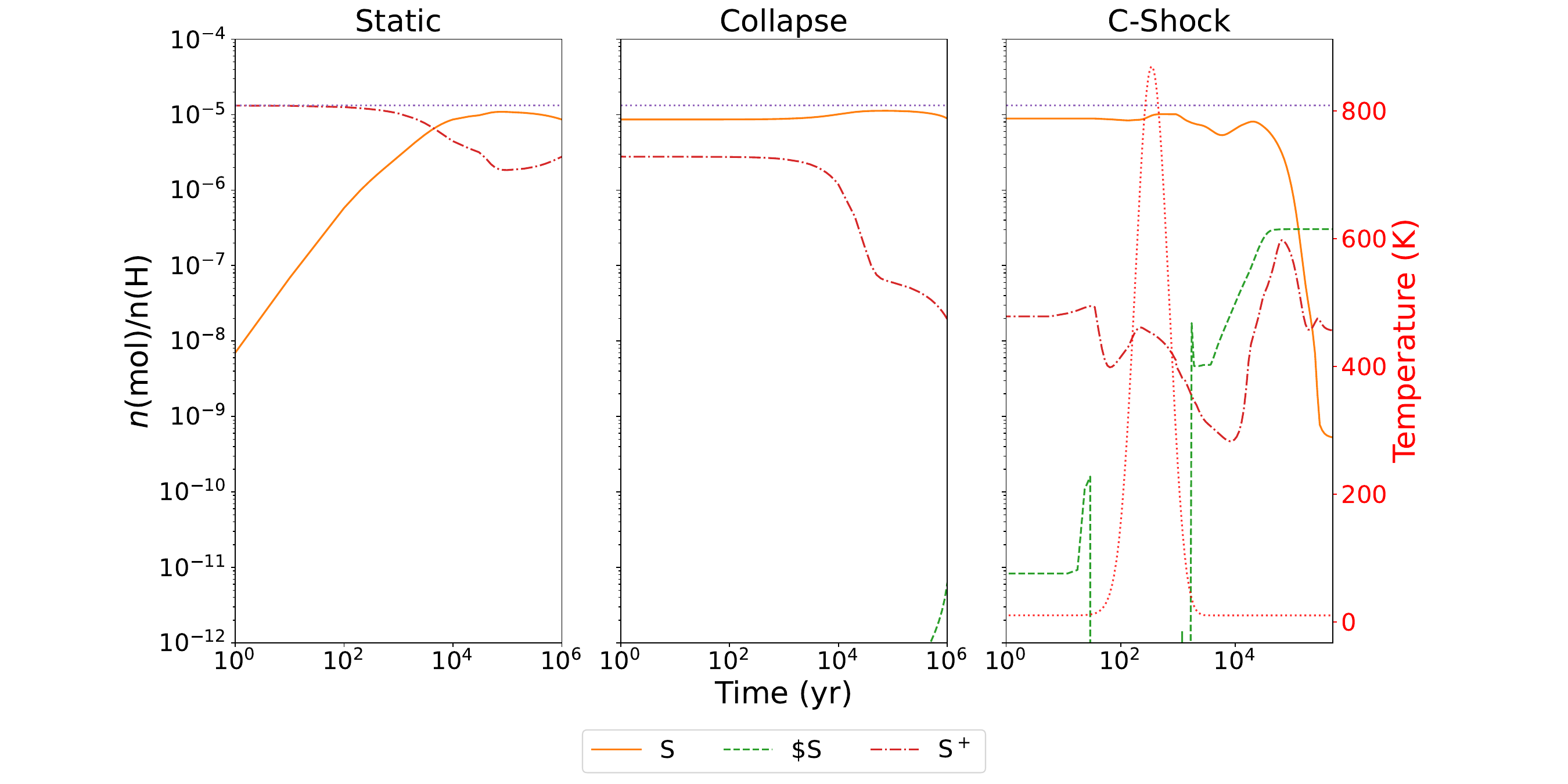}
    \caption{Evolution of S through the different phases of the chemical model. Solid orange line corresponds to gas phase atomic S; red dash-dotted line shows S$^+$ in the gas phase; green dashed line corresponds to S ice in its atomic form. Finally, the horizontal purple dotted line indicates the atomic elemental abundance of S (1.32$\times$10$^{-5}$).}
    \label{fig:s_evol}
\end{figure*}

This abundance is also consistent with the low S depletion factor obtained for highly-irradiated photon-dominated regions \citep{Fuente2023}, which is the expected behaviour for Galactic Center clouds affected by the strong secondary UV photon radiation field produced by the high cosmic-ray ionisation rate.

Figure~\ref{fig:s_evol} shows the evolution of S across the different phases of the chemical model. At the onset of the static phase (phase 0), all S is present in its ionised form. As the model progresses, atomic S begins to form and quickly becomes the dominant sulphur reservoir. It is worth noting that, during the shock propagation the gas‑phase abundance of atomic S reaches approximately 5$\times$10$^{-6}$, only a factor of two below the cosmic S abundance. This is fully consistent with our results for this source.

Thus, from this modelling work, S is likely not depleted towards the Galactic Center molecular cloud G+0.693. This implies that the Galactic Center is not a suitable environment to constrain the amount of solid MgS in dust grains in the ISM. 
However, far-IR observations towards star-forming regions in the Galactic plane could provide some constraints. The {\it PRIMA} space mission will cover several far-IR bands of solid metal sulphides such as MgS and FeS, enabling direct measurements of these metal sulphides in interstellar dust \citep{ PRIMA25}.

\subsection{MgS}
As shown in Section \ref{sec:MgS}, the best fitting model is obtained for a depletion factor of 1000 for Mg. This depletion factor implies that only around 0.1\% of Mg is available in the gas phase, which is sufficient to recover the observed MgS abundances.

\begin{figure*}[h]
    \centering
    \includegraphics[width=0.9\linewidth]{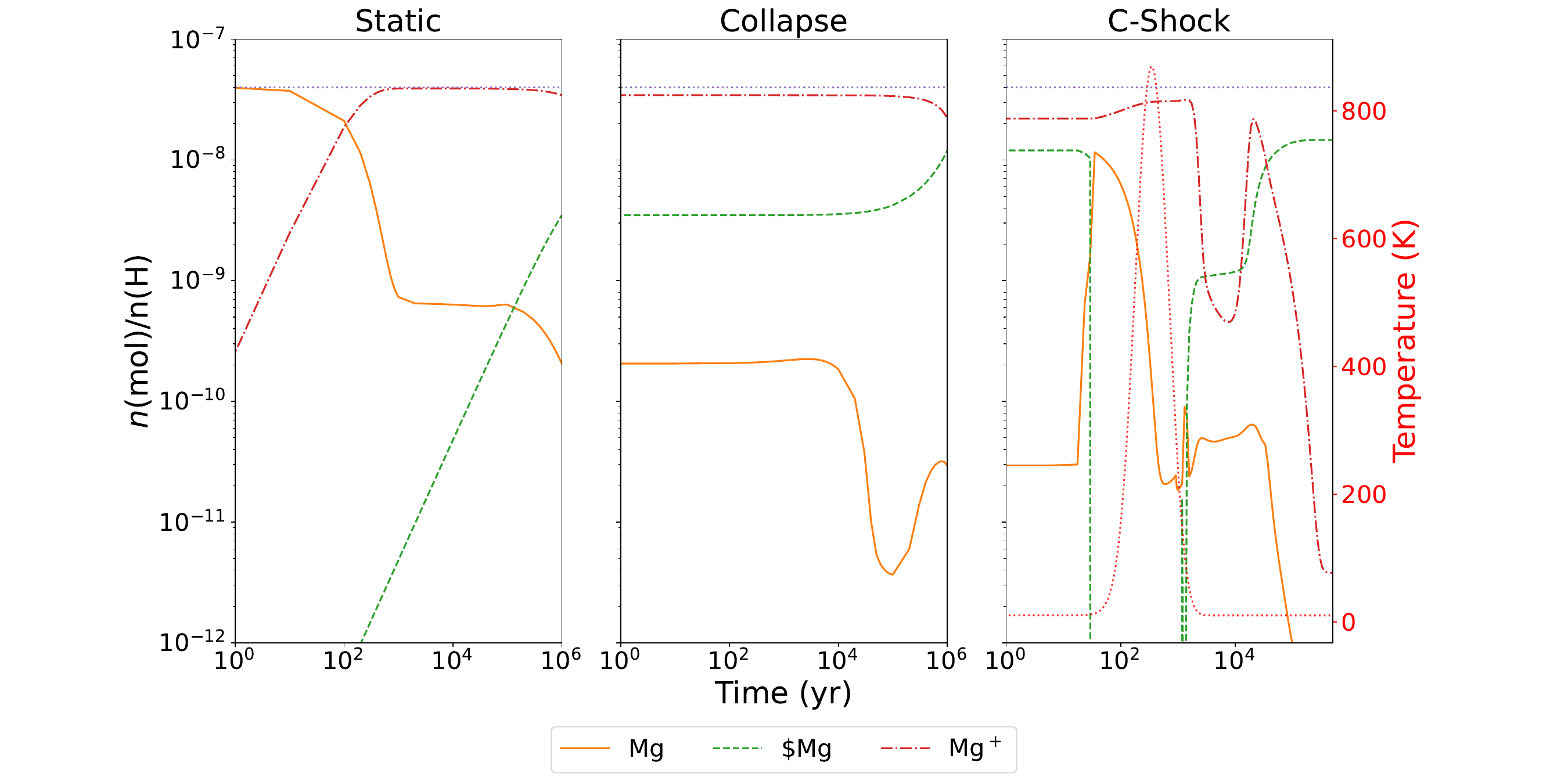}
    \caption{Evolution of Mg through the different phases of the chemical model. Solid orange line corresponds to gas phase atomic Mg; red dash-dotted line shows Mg$^+$ in the gas phase; green dashed line corresponds to ice Mg. Finally, the horizontal purple dotted line shows the atomic elemental abundance of Mg with a depletion factor of 1000 (3.98$\times$10$^{-8}$).}
    \label{fig:mg_evol}
\end{figure*}

Figure~\ref{fig:mg_evol} illustrates the evolution of Mg across the different phases of the chemical model. At the beginning of the static phase, all Mg is present in its atomic form. As the model progresses, Mg becomes increasingly ionised into Mg$^+$. By the end of the collapse phase, roughly one third of the total Mg is locked in ice mantles, while the remaining two thirds reside in Mg$^+$. During the shock phase, atomic Mg is released back into the gas phase as the ices are sputtered, although Mg$^+$ continues being the dominant form.

\begin{figure}[h]
    \centering
    \includegraphics[width=\linewidth]{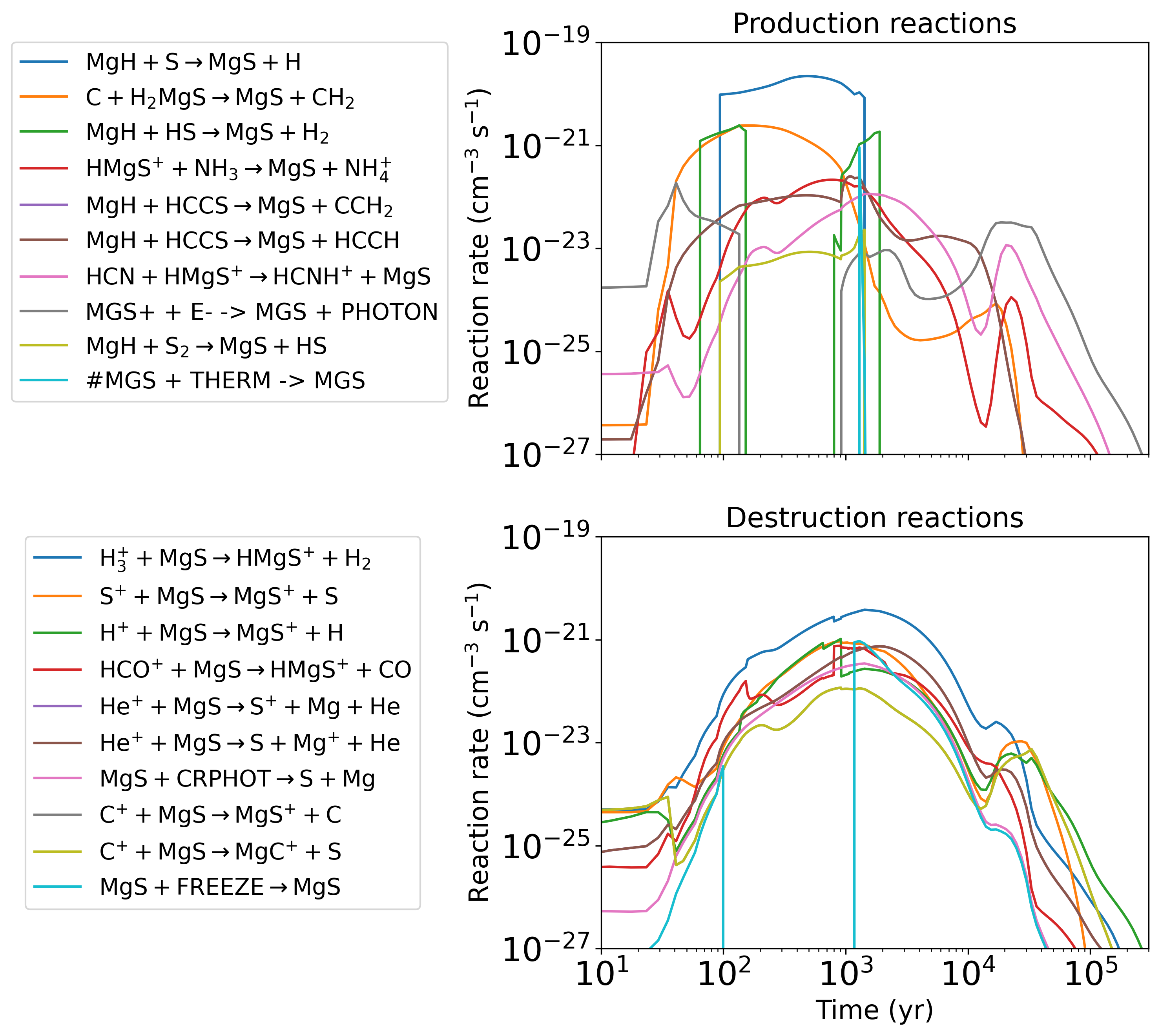}
    \caption{Evolution of the fluxes for the ten dominant production and destruction reactions of MgS during phase 2 within the models with $T_f$=10 K. The legend shows the most significant production (top panel) and destruction (bottom panel) reactions.}
    \label{fig:MGS_analy_reacs}
\end{figure}

Considering the chemical reactions detailed in Section~\ref{subsec:chem}, Figure~\ref{fig:MGS_analy_reacs} presents the evolution of the production and destruction fluxes of MgS through time. We find that most of MgS is produced during the shock through the neutral-neutral (NN) reaction MgH + S $\rightarrow$ MgS + H.

\begin{figure}[h]
    \centering
    \includegraphics[width=\linewidth]{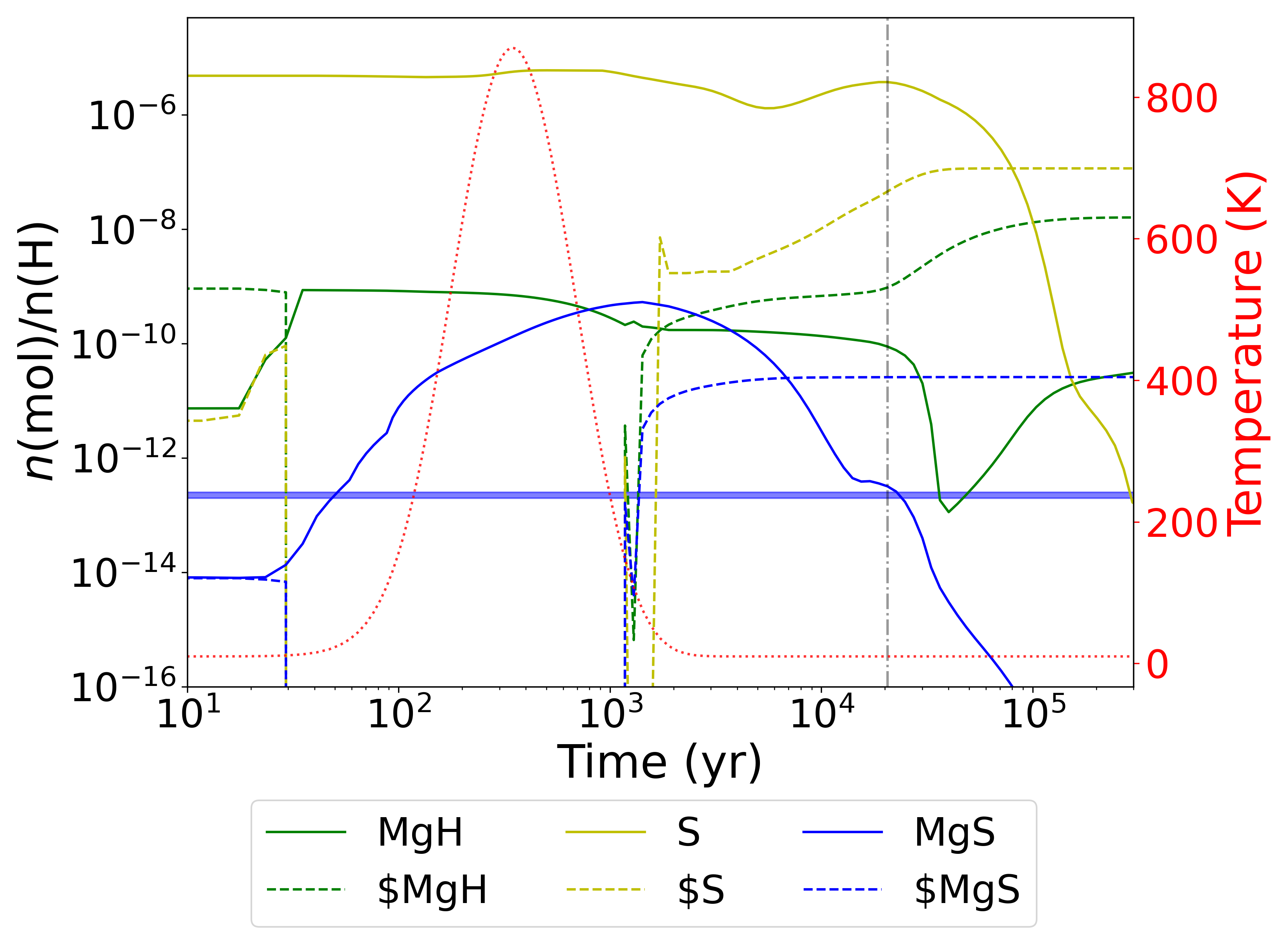}
    \caption{Analysis of MgS (Phase 2). Evolution of the fractional abundance of MgH, S and MgS in the two possible stages: gas phase and ice (surface+bulk; represented by \$). The shaded horizontal line corresponds to the observed abundance of MgS. See caption of Figure~\ref{fig:Deplet_Mg} for more information.}
    \label{fig:MGS_analy}
\end{figure}

Figure~\ref{fig:MGS_analy} presents the evolution of the abundances of MgH, S and MgS in the gas phase and in the ice (surface+bulk; represented by \$) along the propagation of the shock in G+0.693. 
MgH forms through hydrogenation of interstellar ices and it is then released into the gas phase in the shock. Thus, before the shock, MgH is trapped in the icy mantles coating dust grains (see Figure~\ref{fig:MGS_analy}). With the passage of the shock, the ice mantles are sputtered into the gas phase, releasing MgH. Once in the gas phase, MgH can react with S atoms, yielding MgS.

The fact that atomic S is not depleted, meaning that its abundance remains essentially the same across all models with different Mg depletion factors, naturally explains why this reaction becomes so dominant: the reactants remain highly abundant, allowing the reaction to proceed very efficiently. Furthermore, the dominant driver of MgS production is the initial Mg abundance (i.e, the depletion factor) which governs the MgH abundance. 

As the energy dissipates after the shock, the destruction reactions begin to dominate (Figure \ref{fig:MGS_analy_reacs}). For timescales at around 10$^5$ years, the formation of MgS comes from the radiative-recombination (RR) of MgS$^+$ + e$^-$ $\rightarrow$ MgS + $hv$, where MgS$^+$ is formed through the ion-neutral reaction of Mg + HS$^+$ $\rightarrow$ MgS$^+$ + H. The destruction of MgS is dominated by the ion-neutral (IN) reaction H$_3 ^+$ + MgS $\rightarrow$ HMgS$^+$ + H$_2$.\\

Since we are only modelling the chemistry in the gas-phase+ice with UCLCHEM, a high Mg depletion factor of 1000 implies that $\sim$99.9\% of Mg is locked into the core of dust grains and it is not involved in the formation of MgS in the gas phase. This is consistent with the majority of Mg being locked into silicates, a key component of dust grains \citep{Jaeger1994, Dorschener1995, Mutschke1998, Jaeger1998, Fabian2000, Fabian2001, Jaeger2003}\footnote{Some forms of silicates are olivine (Mg$_{2x}$Fe$_{2-2x}$SiO$_4$) or pyroxene (Mg$_x$Fe$_{1-x}$SiO$_3$)}. Consequently, magnesium is likely more present in refractory materials than it will be in the gas phase.

\subsection{Analysis for a final temperature of 100K}

As the energy dissipates in a magnetohydrodynamic shock, colder cores appear \citep[see][]{vanloo2007}. In G+0.693, colder cores with derived gas temperatures down to 30 K have been found \citep{colzi24} within an environment with gas kinetic temperatures between 70 and 150 K \citep{Zeng2018,Zeng2020}. This core is consistent with the cooling of the gas in the postshock regime assumed in our models of Section \ref{sec:results}. However, \citet{willis2025} modelled the chemistry of G+0.693 by assuming the passage of a shock with similar physical parameters to the ones considered here but assuming that the post-shocked gas after the cooling would remain at a temperature higher (100 K) than considered here. Therefore, we complemented our analysis by running additional models assuming a final temperature of 100 K and comparing them to the results obtained in previous sections.\\

\begin{figure}[h]
    \centering
    \includegraphics[width=\linewidth]{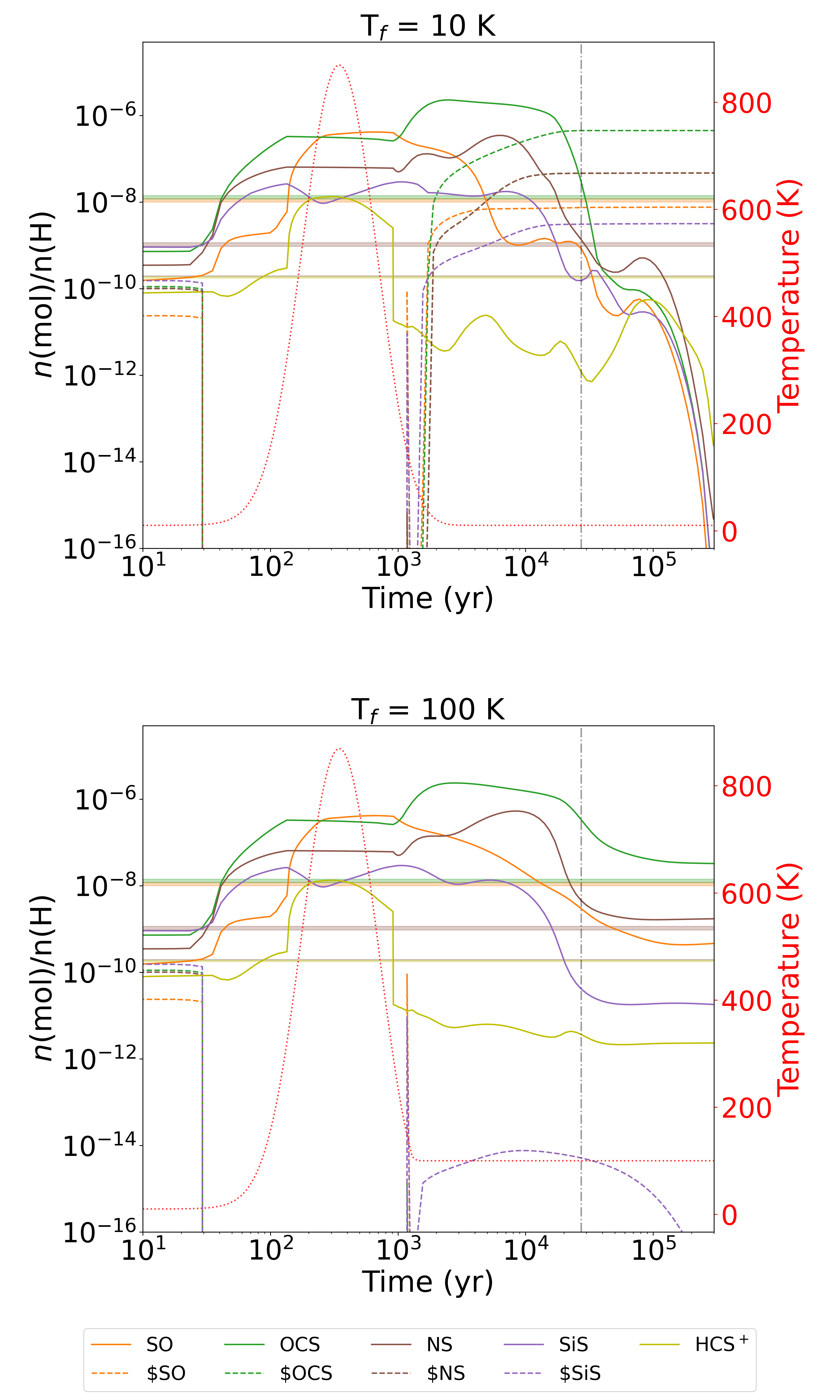}
    \caption{Chemical model results (Phase 2): evolution of the fractional abundances of SO, OCS, NS, SiS and HCS$^+$ in the two possible phases, gas phase and ice (surface+bulk; represented by \$), as a function of time for a depletion factor of 1 for S and a cosmic-ray ionisation rate of $\zeta$=1.3$\times$10$^{-15}$. Top panel corresponds to the results of the analysis assuming a final temperature T$_f$=10 K, whereas the bottom panel shows the results for the model with a final temperature of T$_f$=100 K. See caption of Figure~\ref{fig:S_constrain} for more information.}
    \label{fig:S_comparison}
\end{figure}

From Section~\ref{sec:results}, the results for a final shock temperature of T$_f$=10 K show that S is not depleted and that the cosmic‑ray ionisation rate is enhanced by a factor of 100. Figure~\ref{fig:S_comparison} compares these results with the model computed at T$_f$=100 K using the same S depletion and $\zeta$ parameter. 

When comparing the two models, the main distinction is the predicted final abundances. In the model with a final temperature of 10 K, the abundances decrease by around 10$^5$ years because the molecules begin to freeze out onto the ice mantles (as indicated by the dashed lines in Figure~\ref{fig:S_comparison}). In contrast, the model with T$_f$=100 K does not reproduce the observations because the predicted abundances are overestimated. This is due to the higher temperatures, which prevents the molecules from adhering to the dust grains, and helping them in remaining in the gas phase for longer periods of time.\\

\begin{figure}[h]
    \centering
    \includegraphics[width=\linewidth]{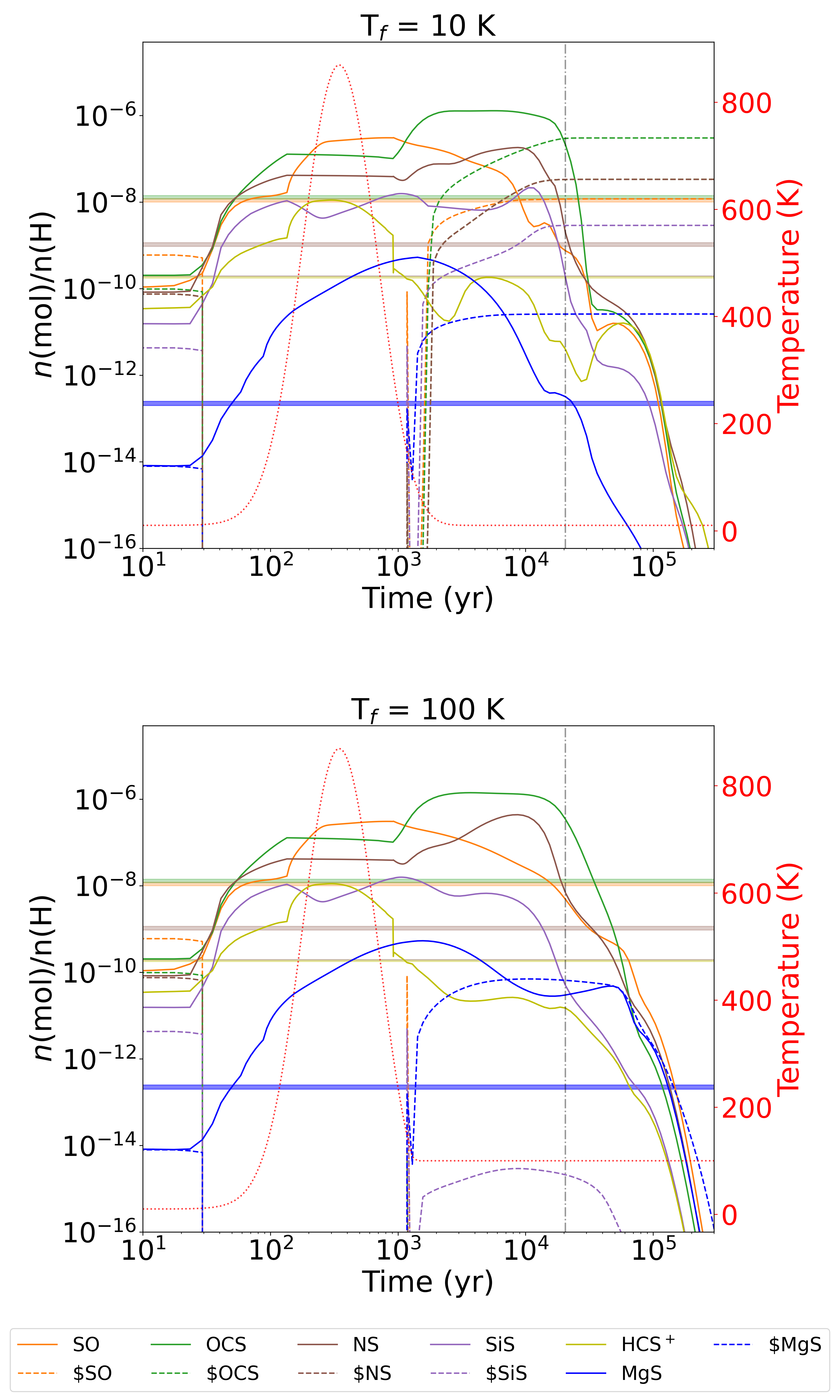}
    \caption{Chemical model results (Phase 2): evolution of the fractional abundances of SO, OCS, NS, SiS, HCS$^+$ and MgS in the two possible stages: gas phase and ice (surface+bulk; represented by \$), as a function of time for a depletion factor of 1000 for Mg and a cosmic-ray ionisation rate of $\zeta$=1.3$\times$10$^{-15}$. Top panel corresponds to the results of the analysis for T$_f$=10 K, whereas bottom panel shows the results for the model with T$_f$=100 K. See caption of Figure~\ref{fig:Deplet_Mg} for more information.}
    \label{fig:Mg_comparison}
\end{figure}

Regarding the MgS analysis, Figure~\ref{fig:Mg_comparison} shows that, for a Mg depletion factor of 1000, the gas phase abundance of MgS at some 10$^4$ years (i.e. similar to the one derived in Section~\ref{sec:MgS}), is approximately three orders of magnitude higher in the model with T$_f$=100 K than in the model with T$_f$=10 K. Since the observed abundance of MgS is of 2.25$\times10^{-12}$, the model with T$_f$=10 K matches better than the model with T$_f$=100 K.

\begin{figure}[h]
    \centering
    \includegraphics[width=\linewidth]{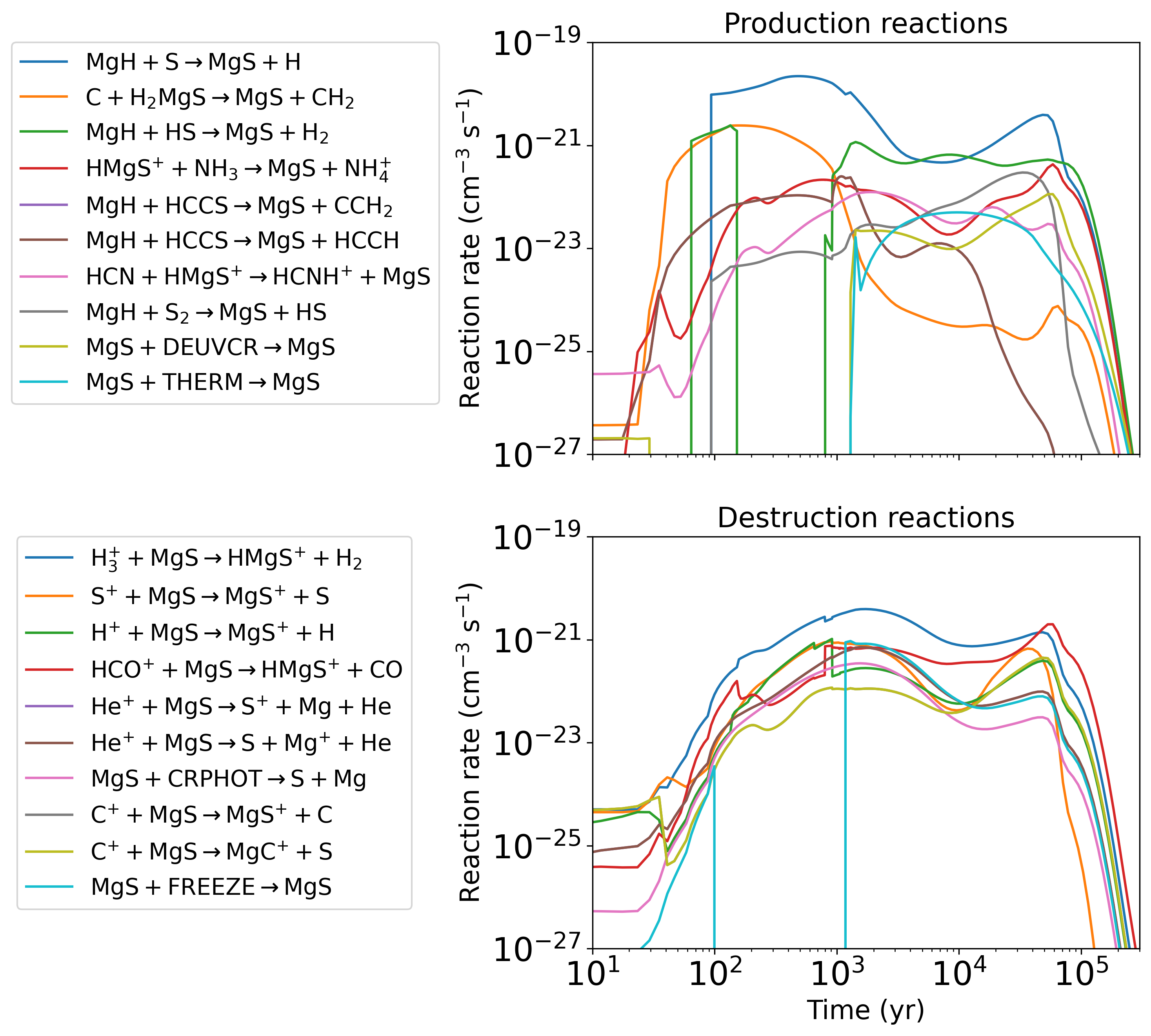}
    \caption{Evolution of the fluxes for the ten dominant production and destruction reactions of MgS during phase 2 within the models with $T_f$=100 K. The legend shows the most significant production (top panel) and destruction (bottom panel) reactions.}
    \label{fig:MGS_analy_reacs_Tf_100}
\end{figure}

The higher modelled abundance for MgS can be understood from Figure~\ref{fig:MGS_analy_reacs_Tf_100} which shows that the production reaction of MgH + S $\rightarrow$ MgS + H competes with the destruction pathways over longer timescales in the T$_f$=100 K model. As a result, the production of MgS is sustained far downstream in the postshock regime, yielding a higher MgS gas‑phase abundance.  
Figure~\ref{fig:MGS_analy_reacs_Tf_100} indicates that the dominant pathways remain unchanged with respect to the T$_f$=10 K model. In both cases, the NN reaction of MgH + S $\rightarrow$ MgS + H is the primary contributor (see Table~\ref{tab:MgS_Tf_100} in Appendix~\ref{sec:mg_analy}).

\subsection{Derived chemical timescale for G+0.693}

In this work, we have derived a chemical\footnote{We refer to chemical timescales not dynamical ones as the passage of the shock resets the chemistry.} timescale of around 2$\times$10$^4$ years, by comparing the UCLCHEM fractional abundances with those observed in G+0.693. 
As illustrated in Figure~\ref{fig:Deplet_Mg}, for timescales around 10$^5$ years the predicted abundances significantly drop and consequently, they largely differ from the observed abundances by several orders of magnitude.

\cite{Requena2006} estimated that recurrent low-velocity shocks with shock velocities of 20 km s$^{-1}$ are needed to explain the constant COM abundances measured across the Galactic Center. The inferred recurring timescales are 5$\times$10$^{4}$ years \citep{Requena2006}, which help sustain the observed large gas-phase abundances of COMs against freeze-out and UV photo-dissociation, becoming efficient at 10$^4$ years after the passage of the shock due to the low temperatures of the dust in the Galactic Center \citep{Etxaluze2013, Battersby2025}. The timescales derived by \cite{Requena2006} are a factor of a few longer than inferred from our models, and are hence broadly consistent. Note however that the simulations of \cite{Requena2006} used a simple analytical model that considered simply the UV photo-destruction of COMs and their recurrent release from ice grains by recurring shocks of timescales of 5$\times$10$^{4}$ years.

More recent studies have demonstrated that, in shock models, most molecules reach their peak abundances shortly after the shock begins, at around 10$^2$ years \citep{willis2025}. Once in the gas phase, these COMs are rapidly destroyed by ion–molecule reactions with abundant ions (in particular, C$^+$ and H$_3^+$) and by cosmic-ray induced photodissociation, which is enhanced by the high cosmic-ray ionisation rate present in the Galactic Center. The main trend identified by \cite{willis2025} from the analysis of shock models with varying cosmic-ray ionisation rates is that higher ionisation values lead to faster gas-phase destruction through ion–molecule reactions and increased cosmic-ray induced photodissociation. At later stages in all shock models, the abundances of most COMs decrease to levels even lower than those prior to the shock due to the elevated gas density, which allows more rapid accretion onto the grains. As a result of the fast depletion onto dust grains, \cite{willis2025} derive timescales of 10$^4$ years for the shock in G+0.693, which agrees with the timescales derived from our analysis.

\section{Conclusions}\label{sec:conclusion}
We have carried out the first chemical modelling of the metal-sulphide MgS towards the Galactic Center cloud G+0.693. We have used the chemical code UCLCHEM to model the chemistry of this species simulating the extreme physical conditions in this cloud. Our goal is to establish a baseline for identifying the main formation and destruction reactions involved in the chemistry of this metal-sulphide and to determine the depletion factors of S and Mg required to reproduce the abundances of MgS and of other S-bearing species observed in the gas phase. This, in turn, allows us to indirectly infer the composition of dust grains.

The conclusions of our work are the following:
\begin{enumerate}
\item The depletion factor for S in G+0.693 is 1 (i.e. no depletion), which is consistent with previous studies \citep{Rivilla2021b, Miguel2024a}. This implies that, according to these models, no S is locked in the cores of dust grains, and therefore, the Galactic Center is not an appropriate environment for constraining the amount of solid MgS present in interstellar dust grains.

\item The Galactic Center is an extreme environment where the chemistry of the G+0.693 molecular cloud is indeed explained by assuming a low-velocity shock with a shock speed of 20 km s$^{-1}$ and a cosmic-ray ionisation rate enhanced by a factor of 100 (or $\zeta$=1.3$\times$10$^{-15}$ s$^{-1}$).
\item The timescales that best match the observed abundances are $\sim$2$\times$10$^4$ years, which are consistent with other chemical modelling carried out for this cloud \citep{willis2025}.
\item To reproduce the observed abundance of MgS, a Mg depletion factor of 1000 is required, indicating that 99.9\% of this element is locked onto dust grains, likely in the form of silicates. Thus, with only 0.1\% of Mg available in the gas phase, we can recover the observed abundances of MgS.
\item The formation of MgS occurs in the gas phase primarily through the reaction MgH + S. Other formation pathways have minor contributions, but a majority of the total reactions considered have been found to present large energy barriers from quantum chemical computations. Future detection of MgH would provide a critical test of this formation pathway.

\end{enumerate}

This study represents the first analysis of the chemistry of MgS in the ISM. Although a large fraction of the rate constants of the gas-phase reactions considered in our model are educated guesses, the dominant processes involved in the formation of MgS have been studied using high-level quantum chemical computations. However, note that our model does not treat the composition of the grain cores but only of the ices (surface + bulk).
Future observations with far-IR space missions such as PRIMA will observe the mid-IR bands of these solid metal sulphides (at 30 $\mu$m for MgS), providing important constraints on the amount of these metal-bearing species into the cores of dust grains.

\begin{acknowledgements}
    This work is supported by ERC grant OPENS, GA No. 101125858, funded by the European Union. Views and opinions expressed are however those of the author(s) only and do not necessarily reflect those of the European Union or the European Research Council Executive Agency. Neither the European Union nor the granting authority can be held responsible for them. M.R.-M., I.J.-S., and L.C. acknowledge funding from the Spanish National Research Council (CSIC) and the ILINK project SENTINEL (ILINK23017) and the Bilateral project SOULMATE (BIJSP25017 ). I.J-.S, L.C. and V.M.R acknowledge funding from grant PID2022-136814NB-I00 funded by the Spanish Ministry of Science, Innovation and Universities/State Agency of Research MICIU/AEI/ 10.13039/501100011033 and by “ERDF/EU”.
    J.G.d.l.C. also acknowledges support from European Funds for Regional Development and the Autonomous Government of Extremadura (Grant No. GR24020).
    The project that gave rise to these results received the support of a fellowship from the ”la Caixa” Foundation (ID 100010434). The fellowship code is LCF/BQ/PR25/12110012.
    TJM's research at QUB is supported by grant ST/T000198/1 from the STFC.
    RCF acknowledges funding from NSF grant AST-2407815 and computational resources provided by the Mississippi Center for Supercomputer Research.
    M.S.-N. acknowledges an Alexander von Humboldt postdoctoral fellowship from the Alexander von Humboldt Foundation
    V.M.R. also acknowledges the grant CNS2023-144464 funded by MICIU/AEI/10.13039/501100011033 and by “European Union NextGenerationEU/PRTR”.
\end{acknowledgements}

\bibliographystyle{aa}
\bibliography{report}

@article{RM2024,
  author = {Rey-Montejo, M. and Jiménez-Serra, I. and Martín-Pintado, J. and others},
  year = {2024},
  journal = {ApJ},
  volume = {975},
  pages = {174},
  doi = {10.3847/1538-4357/ad736e}
}

@article{Field1974,
  author = {Field, G. B.},
  year = {1974},
  journal = {ApJ},
  volume = {187},
  pages = {453},
  doi = {10.1086/152654}
}

@article{Savage1996,
  author = {Savage, B. D. and Sembach, K. R.},
  year = {1996},
  journal = {ApJ},
  volume = {470},
  pages = {893},
  doi = {10.1086/177919}
}

@article{Savaglio2003,
  author = {Savaglio, S. and Fall, S. M. and Fiore, F.},
  year = {2003},
  journal = {ApJ},
  volume = {585},
  pages = {638},
  doi = {10.1086/346225}
}

@article{Jenkins2009,
  author = {Jenkins, E. B.},
  year = {2009},
  journal = {ApJ},
  volume = {700},
  pages = {1299},
  doi = {10.1088/0004-637X/700/2/1299}
}

@article{DCia2015,
  author = {De Cia, A.},
  year = {2015},
  journal = {A\&A},
  volume = {A97},
  pages = {46},
  doi = {10.1051/0004-6361/201527895}
}

@article{Roman2021,
  author = {Roman-Duval, J. and Jenkins, E. B. and Tchernyshyov, K. and others},
  year = {2021},
  journal = {ApJ},
  volume = {910},
  pages = {95},
  doi = {10.3847/1538-4357/abdeb6}
}

@article{Konstant2023,
  author = {Konstantopoulou, C. and De Cia, A. and Krogager, J.-K. and others},
  year = {2023},
  journal = {A\&A},
  volume = {674},
  pages = {C1},
  doi = {10.1051/0004-6361/202243994e}
}

@article{Konstant2024,
  author = {Konstantopoulou, C. and De Cia, A. and Ledoux, C. and others},
  year = {2024},
  journal = {A\&A},
  volume = {681},
  pages = {A64},
  doi = {10.1051/0004-6361/202347171}
}

@article{Izaskun2020,
  author = {Jiménez-Serra, I. and Martín-Pintado, J. and Rivilla, V. M. and others},
  year = {2020},
  journal = {Astrobiology},
  volume = {20},
  pages = {1048},
  doi = {10.1089/ast.2019.2125}
}

@article{Izaskun2022,
  author = {Jiménez-Serra, I. and Rodríguez-Almeida, L. F. and Martín-Pintado, J. and others},
  year = {2022},
  journal = {A\&A},
  volume = {663},
  pages = {A181},
  doi = {10.1051/0004-6361/202142699}
}

@article{Rivilla2019,
  author = {Rivilla, V. M. and Martín-Pintado, J. and Jiménez-Serra, I. and others},
  year = {2019},
  journal = {MNRAS},
  volume = {483},
  pages = {L114},
  doi = {10.1093/mnrasl/sly228}
}

@article{Rivilla2020,
  author = {Rivilla, V. M. and Martín-Pintado, J. and Jiménez-Serra, I. and others},
  year = {2020},
  journal = {ApJL},
  volume = {899},
  pages = {L28},
  doi = {10.3847/2041-8213/abac55}
}

@article{Rivilla2021a,
  author = {Rivilla, V. M. and Jiménez-Serra, I. and Martín-Pintado, J. and others},
  year = {2021},
  journal = {PNAS},
  volume = {118},
  pages = {e2101314118},
  doi = {10.1073/pnas.2101314118}
}

@article{Rivilla2021b,
  author = {Rivilla, V. M. and Jiménez-Serra, I. and García de la Concepción, J. and others},
  year = {2021},
  journal = {MNRAS},
  volume = {506},
  pages = {L79},
  doi = {10.1093/mnrasl/slab074}
}

@article{Rivilla2022a,
  author = {Rivilla, V. M. and Colzi, L. and Jiménez-Serra, I. and others},
  year = {2022},
  journal = {ApJL},
  volume = {929},
  pages = {L11},
  doi = {10.3847/2041-8213/ac6186}
}

@article{Rivilla2022b,
  author = {Rivilla, V. M. and García De La Concepción, J. and Jiménez-Serra, I. and others},
  year = {2022},
  journal = {Frontiers in Astronomy and Space Sciences},
  volume = {9},
  pages = {829288},
  doi = {10.3389/fspas.2022.829288}
}

@article{Rivilla2023,
  author = {Rivilla, V. M. and Sanz-Novo, M. and Jiménez-Serra, I. and others},
  year = {2023},
  journal = {ApJL},
  volume = {953},
  pages = {L20},
  doi = {10.3847/2041-8213/ace977}
}

@article{Rodriguez2021b,
  author = {Rodríguez-Almeida, L. F. and Rivilla, V. M. and Jiménez-Serra, I. and others},
  year = {2021},
  journal = {A\&A},
  volume = {654},
  pages = {L1},
  doi = {10.1051/0004-6361/202141989}
}

@article{Miguel2023,
  author = {Sanz-Novo, M. and Rivilla, V. M. and Jiménez-Serra, I. and others},
  year = {2023},
  journal = {ApJ},
  volume = {954},
  pages = {3},
  doi = {10.3847/1538-4357/ace523}
}

@article{Miguel2024a,
  author = {Sanz-Novo, M. and Rivilla, V. M. and Jiménez-Serra, I. and others},
  year = {2024},
  journal = {ApJ},
  volume = {965},
  pages = {149},
  doi = {10.3847/1538-4357/ad2c01}
}

@article{Miguel2024b,
  author = {Sanz-Novo, M. and Rivilla, V. M. and Müller, H. S. P. and others},
  year = {2024},
  journal = {ApJL},
  volume = {965},
  pages = {L26},
  doi = {10.3847/2041-8213/ad3945}
}

@article{Zeng2021,
  author = {Zeng, S. and Jiménez-Serra, I. and Rivilla, V. M. and others},
  year = {2021},
  journal = {ApJL},
  volume = {920},
  pages = {L27},
  doi = {10.3847/2041-8213/ac2c7e}
}

@article{Zeng2023,
  author = {Zeng, S. and Rivilla, V. M. and Jiménez-Serra, I. and others},
  year = {2023},
  journal = {MNRAS},
  volume = {523},
  pages = {1448},
  doi = {10.1093/mnras/stad1478}
}

@article{SAndres2024,
  author = {San Andrés, D. and Rivilla, V. M. and Colzi, L. and others},
  year = {2024},
  journal = {ApJ},
  volume = {967},
  pages = {39},
  doi = {10.3847/1538-4357/ad3af3}
}

@article{Zeng2020,
  author = {Zeng, S. and Zhang, Q. and Jiménez-Serra, I. and others},
  year = {2020},
  journal = {MNRAS},
  volume = {497},
  pages = {4896},
  doi = {10.1093/mnras/staa2187}
}

@article{Kimura2005,
  author = {Kimura, Y. and Kurumada, M. and Tamura, K. and others},
  year = {2005},
  journal = {A\&A},
  volume = {442},
  pages = {507},
  doi = {10.1051/0004-6361:20052757}
}

@article{Goebel1985,
  author = {Goebel, J. H. and Moseley, S. H.},
  year = {1985},
  journal = {ApJL},
  volume = {290},
  pages = {L35},
  doi = {10.1086/184437}
}

@article{Cernicharo2023,
  author = {Cernicharo, J. and Cabezas, C. and Pardo, J. R. and others},
  year = {2023},
  journal = {A\&A},
  volume = {672},
  pages = {L13},
  doi = {10.1051/0004-6361/202346467}
}

@article{Holdship2017,
  author = {Holdship, J. and Viti, S. and Jiménez-Serra, I. and others},
  year = {2017},
  journal = {AJ},
  volume = {154},
  pages = {38},
  doi = {10.3847/1538-3881/aa773f}
}

@article{Fabian2001,
  author = {Fabian, D. and Henning, T. and Jäger, C. and others},
  year = {2001},
  journal = {A\&A},
  volume = {378},
  pages = {228},
  doi = {10.1051/0004-6361:20011196}
}

@article{Draine2003,
  author = {Draine, B. T.},
  year = {2003},
  journal = {ARA\&A},
  volume = {41},
  pages = {241},
  doi = {10.1146/annurev.astro.41.011802.094840}
}

@article{Turner1985,
  author = {Turner, B. E. and Steimle, T. C.},
  year = {1985},
  journal = {ApJ},
  volume = {299},
  pages = {956},
  doi = {10.1086/163762}
}

@article{Martin1992,
  author = {Martín-Pintado, J. and Bachiller, R. and Fuente, A.},
  year = {1992},
  journal = {A\&A},
  volume = {254},
  pages = {315}
}

@article{Izaskun2008,
  author = {Jiménez-Serra, I. and Caselli, P. and Martín-Pintado, J. and others},
  year = {2008},
  journal = {A\&A},
  volume = {482},
  pages = {549},
  doi = {10.1051/0004-6361:20078054}
}

@article{Schilke1997,
  author = {Schilke, P. and Walmsley, C. M. and Pineau des Forets, G. and others},
  year = {1997},
  journal = {A\&A},
  volume = {321},
  pages = {293}
}

@article{Gusdorf2008,
  author = {Gusdorf, A. and Cabrit, S. and Flower, D. R. and others},
  year = {2008},
  journal = {A\&A},
  volume = {482},
  pages = {809},
  doi = {10.1051/0004-6361:20078900}
}

@article{Podio2017,
  author = {Podio, L. and Codella, C. and Lefloch, B. and others},
  year = {2017},
  journal = {MNRAS},
  volume = {470},
  pages = {L16},
  doi = {10.1093/mnrasl/slx068}
}

@article{Rosi2018,
  author = {Rosi, M. and Mancini, L. and Skouteris, D. and others},
  year = {2018},
  journal = {Chemical Physics Letters},
  volume = {695},
  pages = {87},
  doi = {10.1016/j.cplett.2018.01.053}
}

@article{Mancini2022,
  author = {Mancini, L. and Trinari, M. and Valença Ferreira de Aragão, E. and others},
  year = {2022},
  journal = {Springer International Publishing},
  doi = {10.1007/978-3-031-10562-3_17}
}

@article{Zanchet2018,
  author = {Zanchet, A. and Roncero, O. and Agúndez, M. and others},
  year = {2018},
  journal = {ApJ},
  volume = {862},
  pages = {38},
  doi = {10.3847/1538-4357/aaccff}
}

@article{Quenard2018,
  author = {Quénard, D. and Jiménez-Serra, I. and Viti, S. and others},
  year = {2018},
  journal = {MNRAS},
  volume = {474},
  pages = {2796},
  doi = {10.1093/mnras/stx2960}
}

@article{Millar2024,
  author = {Millar, T. J. and Walsh, C. and Van de Sande, M. and others},
  year = {2024},
  journal = {A\&A},
  volume = {682},
  pages = {A109},
  doi = {10.1051/0004-6361/202346908}
}

@article{Requena2006,
  author = {Requena-Torres, M. A. and Martín-Pintado, J. and Rodríguez-Franco, A. and others},
  year = {2006},
  journal = {A\&A},
  volume = {455},
  pages = {971},
  doi = {10.1051/0004-6361:20065190}
}

@article{Zeng2018,
  author = {Zeng, S. and Jiménez-Serra, I. and Rivilla, V. M. and others},
  year = {2018},
  journal = {MNRAS},
  volume = {478},
  pages = {2962},
  doi = {10.1093/mnras/sty1174}
}

@ARTICLE{Asplund2009,
  author = {Asplund, Martin and Grevesse, Nicolas and Sauval, A. Jacques and Scott, Pat},
  title = "{The Chemical Composition of the Sun}",
  journal = {\araa},
  year = 2009,
  month = sep,
  volume = {47},
  number = {1},
  pages = {481-522},
  doi = {10.1146/annurev.astro.46.060407.145222}
}

@ARTICLE{Izaskun2025,
       author = {{Jim{\'e}nez-Serra}, Izaskun and {Meg{\'\i}as}, Andr{\'e}s and {Salaris}, Joseph and {Cuppen}, Herma and {Taillard}, Ang{\`e}le and {Jin}, Miwha and {Wakelam}, Valentine and {Vasyunin}, Anton I. and {Caselli}, Paola and {Pendleton}, Yvonne J. and {Dartois}, Emmanuel and {Noble}, Jennifer A. and {Viti}, Serena and {Borshcheva}, Katerina and {Garrod}, Robin T. and {Lamberts}, Thanja and {Fraser}, Helen and {Melnick}, Gary and {McClure}, Melissa and {Rocha}, Will and {Drozdovskaya}, Maria N. and {Lis}, Dariusz C.},
        title = "{Modelling methanol and hydride formation in the JWST Ice Age era}",
      journal = {\aap},
         year = 2025,
        month = mar,
       volume = {695},
          eid = {A247},
        pages = {A247},
          doi = {10.1051/0004-6361/202452389},
archivePrefix = {arXiv},
       eprint = {2502.10123},
 primaryClass = {astro-ph.GA},
       adsurl = {https://ui.adsabs.harvard.edu/abs/2025A&A...695A.247J}
}

@article{RodriguezMartin2005,
  author = {Rodríguez-Fernández, N. J. and Martín-Pintado, J.},
  year = {2005},
  journal = {A\&A},
  volume = {429},
  pages = {923},
  doi = {10.1051/0004-6361:20047074}
}

@article{Fuente2023,
  author = {Fuente, A. and Riviére-Marichalar, P. and Beitia-Antero, L. and others},
  year = {2023},
  journal = {A\&A},
  volume = {670},
  pages = {A114},
  doi = {10.1051/0004-6361/202244843}
}

@article{Neufeld2015,
  author = {Neufeld, D. A. and Godard, B. and Gerin, M. and others},
  year = {2015},
  journal = {A\&A},
  volume = {577},
  pages = {A49},
  doi = {10.1051/0004-6361/201425391}
}

@article{wei25,
  author = {Wei, Q. and Chen, Y. and Xiao, L. and others},
  year = {2025},
  journal = {A\&A},
  volume = {699},
  pages = {A264},
  doi = {10.1051/0004-6361/202452981}
}

@article{bell25,
  author = {Bell, K. M. and Fortenberry, R. C.},
  year = {2025},
  journal = {Molecules},
  number = {8},
  pages = {1650},
  title = {The Formation of MgS \& MgO Monomers and Dimers from Magnesium, Oxygen, and Sulfur Hydrides},
  doi = {10.3390/molecules30081650}
}

@ARTICLE{willis2025,
       author = {{Willis}, Sydney A. and {Krasnokutski}, Serge A. and {Morin}, Nathaniel J. and {Garrod}, Robin T.},
        title = "{Chemical modeling of aminoketene, ethanolamine, and glycine production in interstellar ices}",
      journal = {\aap},
         year = 2025,
        month = oct,
       volume = {703},
          eid = {A27},
        pages = {A27},
          doi = {10.1051/0004-6361/202554598},
archivePrefix = {arXiv},
       eprint = {2510.20912},
 primaryClass = {astro-ph.GA},
       adsurl = {https://ui.adsabs.harvard.edu/abs/2025A&A...703A..27W}
}

@article{PRIMA25,
  author = {Jiménez-Serra, I. and Zeng, S. and Yang, Y.-L. and others},
  year = {2025},
  journal = {Journal of Astronomical Telescopes, Instruments, and Systems},
  volume = {11},
  pages = {031618},
  doi = {10.1117/1.JATIS.11.3.031618}
}

@article{colzi24,
  author = {Colzi, L. and Martín-Pintado, J. and Zeng, S. and others},
  year = {2024},
  journal = {A\&A},
  volume = {690},
  pages = {A121},
  doi = {10.1051/0004-6361/202451382}
}

@article{Pechukas1965,
  author = {Pechukas, P. and Light, J. C.},
  year = {1965},
  journal = {J. Chem. Phys.},
  volume = {42},
  pages = {3281},
  doi = {10.1063/1.1696411}
}

@article{Chesnavich1986,
  author = {Chesnavich, W. J.},
  year = {1986},
  journal = {J. Chem. Phys.},
  volume = {84},
  pages = {2615},
  doi = {10.1063/1.450331}
}

@article{Krieger2017,
  author = {Krieger, N. and Ott, J. and Beuther, H. and others},
  year = {2017},
  journal = {ApJ},
  volume = {850},
  pages = {77},
  doi = {10.3847/1538-4357/aa951c}
}

@article{Wakelam2006,
  author = {Wakelam, V. and Herbst, E. and Selsis, F.},
  year = {2006},
  journal = {A\&A},
  volume = {451},
  pages = {551},
  doi = {10.1051/0004-6361:20054682}
}

@article{Jaeger1994,
  author = {Jaeger, C. and Mutschke, H. and Begemann, B. and others},
  year = {1994},
  journal = {A\&A},
  volume = {292},
  pages = {641}
}

@article{Dorschener1995,
  author = {Dorschner, J. and Begemann, B. and Henning, T. and others},
  year = {1995},
  journal = {A\&A},
  volume = {300},
  pages = {503}
}

@article{Mutschke1998,
  author = {Mutschke, H. and Begemann, B. and Dorschner, J. and others},
  year = {1998},
  journal = {A\&A},
  volume = {333},
  pages = {188}
}

@article{Jaeger1998,
  author = {Jaeger, C. and Molster, F. J. and Dorschner, J. and others},
  year = {1998},
  journal = {A\&A},
  volume = {339},
  pages = {904}
}

@article{Fabian2000,
  author = {Fabian, D. and Jäger, C. and Henning, T. and others},
  year = {2000},
  journal = {A\&A},
  volume = {364},
  pages = {282}
}

@article{Jaeger2003,
  author = {Jäger, C. and Dorschner, J. and Mutschke, H. and others},
  year = {2003},
  journal = {A\&A},
  volume = {408},
  pages = {193},
  doi = {10.1051/0004-6361:20030916}
}

@article{vanloo2007,
  author = {van Loo, S. and Falle, S. A. E. G. and Hartquist, T. W. and others},
  year = {2007},
  journal = {A\&A},
  volume = {471},
  pages = {213},
  doi = {10.1051/0004-6361:20077430}
}

@article{Etxaluze2013,
  author = {Etxaluze, M. and Goicoechea, J. R. and Cernicharo, J. and others},
  year = {2013},
  journal = {A\&A},
  volume = {556},
  pages = {A137},
  doi = {10.1051/0004-6361/201321258}
}

@article{Battersby2025,
  author = {Battersby, C. and Walker, D. L. and Barnes, A. and others},
  year = {2025},
  journal = {ApJ},
  volume = {984},
  pages = {156},
  doi = {10.3847/1538-4357/adb5f0}
}

@article{Miguel2025,
  author = {Sanz-Novo, M. and Rivilla, V. M. and Endres, C. P. and others},
  year = {2025},
  journal = {ApJL},
  volume = {980},
  pages = {L37},
  doi = {10.3847/2041-8213/adafa7}
}

@article{Araki2026,
  author = {Araki, M. and Sanz-Novo, M. and Endres, C. P. and others},
  year = {2026},
  journal = {Nature Astronomy},
  volume = {10},
  pages = {401},
  doi = {10.1038/s41550-025-02749-7}
}

@article{Georgievskii2013,
  author = {Georgievskii, Y. and Miller, J. A. and Burke, M. P. and others},
  year = {2013},
  journal = {J. Phys. Chem. A},
  volume = {117},
  pages = {12146},
  doi = {10.1021/jp4060704}
}

@article{colzi22,
  author = {Colzi, L. and Martín-Pintado, J. and Rivilla, V. M. and others},
  year = {2022},
  journal = {ApJL},
  volume = {926},
  pages = {L22},
  doi = {10.3847/2041-8213/ac52ac}
}

@article{Massalkhi2023,
  author = {Massalkhi, S. and Jiménez-Serra, I. and Martín-Pintado, J. and others},
  year = {2023},
  journal = {A\&A},
  volume = {678},
  pages = {A45},
  doi = {10.1051/0004-6361/202346822}
}

@article{martin2008,
  author = {Martín, S. and Requena-Torres, M. A. and Martín-Pintado, J. and others},
  year = {2008},
  journal = {ApJ},
  volume = {678},
  pages = {245},
  doi = {10.1086/533409}
}

@article{Fortenberry2024,
  author = {Fortenberry, R. C. and McGuire, B. A.},
  year = {2024},
  journal = {ApJ},
  volume = {971},
  pages = {101},
  doi = {10.3847/1538-4357/ad4d94}
}

@article{Adler2007,
  author = {Adler, T. B. and Knizia, G. and Werner, H.-J.},
  year = {2007},
  journal = {J. Chem. Phys.},
  volume = {127},
  pages = {221106},
  doi = {10.1063/1.2817618}
}

@article{Knizia2009,
  author = {Knizia, G. and Adler, T. B. and Werner, H.-J.},
  year = {2009},
  journal = {J. Chem. Phys.},
  volume = {130},
  pages = {054104},
  doi = {10.1063/1.3054300}
}

@article{Peterson2008,
  author = {Peterson, K. A. and Adler, T. B. and Werner, H.-J.},
  year = {2008},
  journal = {J. Chem. Phys.},
  volume = {128},
  pages = {084102},
  doi = {10.1063/1.2831537}
}

@article{Werner2012,
  author = {Werner, H.-J. and Knowles, P. J. and Knizia, G. and others},
  year = {2012},
  journal = {WIREs Comput. Mol. Sci.},
  volume = {2},
  pages = {242},
  doi = {10.1002/wcms.82}
}

@ARTICLE{Rivilla2026,
       author = {{Rivilla}, V.~M. and {San Andr{\'e}s}, D. and {Sanz-Novo}, M. and {Colzi}, L. and {Jim{\'e}nez-Serra}, I. and {L{\'o}pez-Gallifa}, A. and {Mart{\'\i}nez-Henares}, A. and {Meg{\'\i}as}, A. and {Mart{\'\i}n}, S. and {Tercero}, B. and {Zeng}, S. and {Loreau}, J. and {Ben Khalifa}, M. and {Requena-Torres}, M.~A. and {de Vicente}, P.},
        title = "{Aromatic rings in the Central Molecular Zone: Benzonitrile}",
      journal = {A\&A},
         year = 2026,
        month = apr,
          eid = {arXiv:2604.24510},
        pages = {arXiv:2604.24510},
          doi = {10.48550/arXiv.2604.24510},
archivePrefix = {arXiv},
       eprint = {2604.24510},
 primaryClass = {astro-ph.GA},
       adsurl = {https://ui.adsabs.harvard.edu/abs/2026arXiv260424510R}
}

@ARTICLE{Andres2025,
       author = {{Meg{\'\i}as}, Andr{\'e}s and {Jim{\'e}nez-Serra}, Izaskun and {Dulieu}, Fran{\c{c}}ois and {Vitorino}, Julie and {Mat{\'e}}, Bel{\'e}n and {Ciudad}, David and {Rocha}, Will R.~M. and {Mart{\'\i}nez Jim{\'e}nez}, Marcos and {Aguirre}, Jacobo},
        title = "{A fast machine learning tool to predict the composition of interstellar ices from infrared absorption spectra}",
      journal = {\aap},
         year = 2025,
        month = oct,
       volume = {702},
          eid = {A87},
        pages = {A87},
          doi = {10.1051/0004-6361/202453640},
archivePrefix = {arXiv},
       eprint = {2509.04331},
 primaryClass = {astro-ph.GA},
       adsurl = {https://ui.adsabs.harvard.edu/abs/2025A&A...702A..87M}
}

@ARTICLE{Alvaro2024,
       author = {{L{\'o}pez-Gallifa}, {\'A}. and {Rivilla}, V.~M. and {Beltr{\'a}n}, M.~T. and {Colzi}, L. and {Mininni}, C. and {S{\'a}nchez-Monge}, {\'A}. and {Fontani}, F. and {Viti}, S. and {Jim{\'e}nez-Serra}, I. and {Testi}, L. and {Cesaroni}, R. and {Lorenzani}, A.},
        title = "{The GUAPOS project - V: The chemical ingredients of a massive stellar protocluster in the making}",
      journal = {\mnras},
         year = 2024,
        month = apr,
       volume = {529},
       number = {4},
        pages = {3244-3283},
          doi = {10.1093/mnras/stae676},
archivePrefix = {arXiv},
       eprint = {2403.02191},
 primaryClass = {astro-ph.GA},
       adsurl = {https://ui.adsabs.harvard.edu/abs/2024MNRAS.529.3244L}
}

@ARTICLE{Campanha2022,
       author = {{Campanha}, Danilo R. and {Mendoza}, Edgar and {Silva}, Mateus X. and {Velloso}, Paulo F.~G. and {Carvajal}, Miguel and {Wakelam}, Valentine and {Galv{\~a}o}, Breno R.~L.},
        title = "{The Si + SO$_{2}$ collision and an extended network of neutral-neutral reactions between silicon and sulphur bearing species}",
      journal = {\mnras},
         year = 2022,
        month = sep,
       volume = {515},
       number = {1},
        pages = {369-377},
          doi = {10.1093/mnras/stac1647},
archivePrefix = {arXiv},
       eprint = {2206.04982},
 primaryClass = {astro-ph.GA},
       adsurl = {https://ui.adsabs.harvard.edu/abs/2022MNRAS.515..369C}
}

\onecolumn
\begin{appendix} \label{sec:appendix}
\renewcommand{\arraystretch}{1.2} 
\section{Updated chemical network}
In this section, we present the chemical network that was added to the RATE22 database for gas-phase reactions (Tables~\ref{tab:chemical_reactions_SIS} and~\ref{tab:reac_MGS}) as well as to the default dust-grain network for grain-surface reactions (Tables~\ref{tab:chem_mgs_grain}). Moreover, in Table~\ref{tab:new_reac_MGS}, we show the new reactions studied and included in the chemical network.

\subsection{SiS}
\begin{table*}[h]
    \centering
    \caption{Gas-phase chemical reactions added for SiS. The coefficients $\alpha$, $\beta$, $\gamma$ follow the Kooij-Arrhenius equation $k(\rm T) = \alpha (\frac{T}{300K})^{\beta}\rm exp (-\gamma/T)$, where T corresponds to the temperature range included in the last column.}
    \begin{tabular}{l c c c c c} 
        \toprule
        {Reaction} & {Type} & {$\alpha$} & {$\beta$} & {$\gamma$} & {Temperature range (K)}\\
        \midrule
        \ch{SiH$_2$ + S$^+$ -> SiSH$^+$ + H} & IN & $1.00\times10^{-09}$ & 0.0 & 0.0 & 10:41000\\
        \ch{SiH$_2$ + S$^+$ -> HSiS$^+$ + H} & IN & $1.00\times10^{-09}$ & 0.0 & 0.0 & 10:41000\\
        \ch{HSiS$^+$ + NH$_3$ -> SiS + NH$_4$$^+$} & IN & $1.00\times10^{-09}$ & 0.0 & 0.0 & 10:41000\\
        \bottomrule
    \end{tabular}
    \label{tab:chemical_reactions_SIS}
\begin{tablenotes}
\footnotesize
\item Note 1. Reactions from \cite{Mancini2022}. $\alpha$, $\beta$ and $\gamma$ parameters are educated guesses.
\item Note 2. Type reactions account for: IN (ion-neutral).
\end{tablenotes}
\end{table*}

\subsection{MgS}
\begin{ThreePartTable}
\begin{longtable}{l c c c c c}
\caption{Gas-phase chemical reactions added for MgS. The coefficients $\alpha$, $\beta$, $\gamma$ follow the Kooij-Arrhenius equation $k(\rm T) = \alpha (\frac{T}{300K})^{\beta}\rm exp (-\gamma/T)$, where T corresponds to the temperature range included in the last column. }

\\
    \toprule
    {Reaction} & {Type} & {$\alpha$} & {$\beta$} & {$\gamma$} & {Temperature range (K)}\\
    \midrule
    \endfirsthead

    \multicolumn{6}{c}{\textit{continued from previous page}} \\
    \toprule
    {Reaction} & {Type} & {$\alpha$} & {$\beta$} & {$\gamma$} & {Temperature range (K)}\\
    \midrule
    \endhead

    \midrule \multicolumn{6}{r}{\textit{Continued on next page}} \\
    \endfoot

    \bottomrule
    \endlastfoot
        $^*$\ch{C$^+$ + MgS -> MgS$^+$ + C} & CE & $4.35\times10^{-09}$ & -0.50 & 0.0 & 10:41000\\
        $^*$\ch{H$^+$ + MgS -> MgS$^+$ + H} & CE & $2.76\times10^{-08}$ & -0.50 & 0.0 & 10:41000\\
        $^*$\ch{S$^+$ + MgS -> MgS$^+$ + S} & CE & $5.46\times10^{-09}$ & -0.50 & 0.0 & 10:41000\\
        \ch{MgS + CRPHOT -> S + Mg} & CR & $1.30\times10^{-17}$ & 0.0 & 250.0 & 10:41000\\
        \ch{MgSH$^+$ + e$^-$ -> MgS + H} & DR & $1.50\times10^{-07}$ & -0.50 & 0.0 & 10:300\\
        $^*$\ch{C$^+$ + MgS -> MgC$^+$ + S} & IN & $4.35\times10^{-09}$ & -0.50 & 0.0 & 10:41000\\
        \ch{H$_2$S + HMgS$^+$ -> H$_3$S$^+$ + MgS} & IN & $2.90\times10^{-10}$ & -0.50 & 0.0 & 10:41000\\
        $^*$\ch{H$_3^+$ + MgS -> HMgS$^+$ + H$_2$} & IN & $1.62\times10^{-08}$ & -0.50 & 0.0 & 10:41000\\
        \ch{HCN + HMgS$^+$ -> HCNH$^+$ + MgS} & IN & $6.10\times10^{-10}$ & -0.50 & 0.0 & 10:41000\\
        $^*$\ch{HCO$^+$ + MgS -> HMgS$^+$ + CO} & IN & $6.26\times10^{-09}$ & -0.50 & 0.0 & 10:41000\\
        $^*$\ch{He$^+$ + MgS -> S$^+$ + Mg + He} & IN & $7.08\times10^{-09}$ & -0.50 & 0.0 & 10:41000\\
        $^*$\ch{He$^+$ + MgS -> S + Mg$^+$ + He} & IN & $7.08\times10^{-09}$ & -0.50 & 0.0 & 10:41000\\
        \ch{NH$_3$ + MgSH$^+$ -> NH$_4 ^+$ + MgS} & IN & $9.70\times10^{-10}$ & -0.50 & 0.0 & 10:41000\\
        \ch{MgS + $hv$ -> S + Mg} & PH & $1.0\times10^{-10}$ & 0.0 & 2.3 & 10:41000\\
        \ch{MgSH$^+$ + e$^-$ -> HS + Mg} & DR & $1.50\times10^{-07}$ & -0.50 & 0.0 & 10:300\\
        $^{****}$\ch{MgS$^+$ + e$^-$ -> S + Mg} & DR & $2.0\times10^{-07}$ & -0.50 & 0.0 & 10:300\\
        \ch{H$_2$O + HMgS$^+$ -> MgOH$^+$ + H$_2$S} & IN & $1.10\times10^{-09}$ & -0.50 & 0.0 & 10:41000\\
        \ch{H + MgS$^+$ -> HS + Mg$^+$} & IN & $1.90\times10^{-09}$ & 0.0 & 0.0 & 10:41000\\
        \ch{O$_2$ + MgS$^+$ -> SO$^+$ + MgO} & IN & $6.23\times10^{-11}$ & 0.0 & 0.0 & 10:41000\\
        \ch{O$_2$ + MgS$^+$ -> MgO$^+$ + SO} & IN & $2.67\times10^{-11}$ & 0.0 & 0.0 & 10:41000\\
        $^{****}$\ch{Mg$^+$ + OCS -> MgS$^+$ + CO} & IN & 0.0 & 0.0 & 0.0 & 10:41000\\
        \ch{MgH$_2^+$ + S -> HMgS$^+$ + H} & IN & $1.10\times10^{-09}$ & 0.0 & 0.0 & 10:41000\\
        $^*$\ch{MgH + S$^+$ -> MgS$^+$ + H} & IN & $1.37\times10^{-09}$ & 0.0 & 0.0 & 10:41000\\
        \ch{HMgS + CRPHOT -> MgS + H} & CR & $1.30\times10^{-17}$ & 0.0 & 250.0 & 10:41000\\
        \ch{MgSH$_2$ + CRPHOT -> HMgS + H} & CR & $1.30\times10^{-17}$ & 0.0 & 125.0 & 10:41000\\
        \ch{MgSH$_2$ + CRPHOT -> MgS + H$_2$} & CR & $1.30\times10^{-17}$ & 0.0 & 125.0 & 10:41000\\
        \ch{MgSH$_2$$^+$ + e$^-$ -> MgS + H$_2$} & DR & $7.50\times10^{-08}$ & -0.50 & 0.0 & 10:300\\
        \ch{MgSH$_2$$^+$ + e$^-$ -> HMgS + H} & DR & $7.50\times10^{-08}$ & -0.50 & 0.0 & 10:300\\
        \ch{H$_3$MgS$^+$ + e$^-$ -> HMgS + H$_2$} & DR & $7.50\times10^{-08}$ & -0.50 & 0.0 & 10:300\\
        \ch{H$_3$MgS$^+$ + e$^-$ -> MgSH$_2$ + H} & DR & $7.50\times10^{-08}$ & -0.50 & 0.0 & 10:300\\
         $^*$\ch{C$^+$ + HMgS -> MgS + CH$^+$} & IN & $3.64\times10^{-09}$ & -0.50 & 0.0 & 10:41000\\
         $^*$\ch{H$_3$$^+$ + HMgS -> MgSH$_2$$^+$ + H$_2$} & IN & $6.76\times10^{-09}$ & -0.50 & 0.0 & 10:41000\\
        \ch{HCN + MgSH$_2$$^+$ -> HCNH$^+$ + HMgS} & IN & $3.05\times10^{-10}$ & -0.50 & 0.0 & 10:41000\\
         $^*$\ch{HCO$^+$ + HMgS -> MgSH$_2$$^+$ + CO} & IN & $2.61\times10^{-09}$ & -0.50 & 0.0 & 10:41000\\
         $^*$\ch{He$^+$ + HMgS -> MgS$^+$ + H + He} & IN & $5.93\times10^{-09}$ & -0.50 & 0.0 & 10:41000\\
        \ch{NH$_3$ + MgSH$_2$$^+$ -> NH$_4$$^+$ + HMgS} & IN & $4.85\times10^{-10}$ & -0.50 & 0.0 & 10:41000\\
        \ch{C$^+$ + MgSH$_2$ -> MgS + CH$_2$$^+$} & IN & $3.26\times10^{-09}$ & -0.50 & 0.0 & 10:41000\\
        \ch{H$_3$$^+$ + MgSH$_2$ -> H$_3$MgS$^+$ + H$_2$} & IN & $6.11\times10^{-09}$ & -0.50 & 0.0 & 10:41000\\
        \ch{HCO$^+$ + MgSH$_2$ -> H$_3$MgS$^+$ + CO} & IN & $2.32\times10^{-09}$ & -0.50 & 0.0 & 10:41000\\
        \ch{He$^+$ + MgSH$_2$ -> MgS$^+$ + H$_2$ + He} & IN & $5.33\times10^{-09}$ & -0.50 & 0.0 & 10:41000\\
        \ch{NH$_3$ + H$_3$MgS$^+$ -> NH$_4$$^+$ + MgSH$_2$} & IN & $4.85\times10^{-10}$ & -0.50 & 0.0 & 10:41000\\
        \ch{HCN + H$_3$MgS$^+$ -> HCNH$^+$ + MgSH$_2$} & IN & $3.05\times10^{-10}$ & -0.50 & 0.0 & 10:41000\\
        \ch{C + HMgS -> MgS + CH} & NN & $2.0\times10^{-10}$ & 0.0 & 0.0 & 10:41000\\
        \ch{H + HMgS -> MgS + H$_2$} & NN & $1.0\times10^{-11}$ & 0.0 & 0.0 & 10:41000\\
        \ch{OH + HMgS -> MgS + H$_2$O} & NN & $2.0\times10^{-10}$ & 0.0 & 0.0 & 10:41000\\
        $^{****}$\ch{Mg + HS -> MgS + H} & NN & 0.0 & 0.0 & 0.0 & 10:41000\\
        $^{****}$\ch{Mg + H$_2$S -> MgS + H$_2$} & NN & 0.0 & 0.0 & 0.0 & 10:300\\
        $^{****}$\ch{Mg + H$_2$S -> HMgS + H} & NN & 0.0 & 0.0 & 0.0 & 10:300\\
        $^*$\ch{O + MgS -> MgO + S} & NN & $9.53\times10^{-11}$ & 0.29 & 6000 & 50:2500\\
        \ch{O + HMgS -> MgS + OH} & NN & $2.0\times10^{-10}$ & 0.0 & 0.0 & 10:41000\\
        $^*$\ch{Mg + SO -> MgS + O} & NN & $0.0$ & 0.16 & -20.0 & 100:2500\\
        $^*$\ch{Mg + SO$_2$ -> MgS + O$_2$} & NN & $0.0$ & 0.0 & 0.0 & 100:2500\\
        \ch{MgH + S -> MgS + H} & NN & $1.0\times10^{-10}$ & 0.0 & 0.0 & 100:2500\\
        \ch{MgH + S$_2$ -> MgS + HS} & NN & $1.0\times10^{-10}$ & 0.0 & 0.0 & 100:2500\\
        \ch{MgH + H$_2$S -> MgSH$_2$ + H} & NN & $3.0\times10^{-10}$ & 0.0 & 0.0 & 10:300\\
        \ch{MgH + H$_2$S -> HMgS + H$_2$} & NN & $3.0\times10^{-10}$ & 0.0 & 0.0 & 10:300\\
        \ch{O + MgSH$_2$ -> HMgS + OH} & NN & $2.00\times10^{-10}$ & 0.0 & 0.0 & 10:41000\\
        \ch{C + MgSH$_2$ -> MgS + CH$_2$} & NN & $2.0\times10^{-10}$ & 0.0 & 0.0 & 10:41000\\
        \ch{OH + MgSH$_2$ -> HMgS + H$_2$O} & NN & $2.0\times10^{-10}$ & 0.0 & 0.0 & 10:41000\\
        \ch{HMgS + $hv$ -> H + MgS} & PH & $1.00\times10^{-10}$ & 0.0 & 2.3 & 10:41000\\
        \ch{MgSH$_2$ + $hv$ -> H + HMgS} & PH & $5.0\times10^{-11}$ & 0.0 & 2.3 & 10:41000\\
        \ch{MgSH$_2$ + $hv$ -> H$_2$ + MgS} & PH & $5.0\times10^{-11}$ & 0.0 & 2.3 & 10:41000\\
        $^{***}$\ch{Mg + S -> MgS + $hv$} & RA & $6.627\times10^{-18}$ & 0.0864 & 2.7057 & 10:50\\
        $^{***}$\ch{Mg + S -> MgS + $hv$} & RA & $4.7972\times10^{-18}$ & -0.3143 & 22.7454 & 50:100\\
        $^{***}$\ch{Mg + S -> MgS + $hv$} & RA & $4.5528\times10^{-18}$ & 0.1682 & -39.665 & 100:800\\
        $^{***}$\ch{Mg + S -> MgS + $hv$} & RA & $1.8002\times10^{-19}$ & 1.939 & -1258.5349 & 800:2000\\
        $^{***}$\ch{Mg + S -> MgS + $hv$} & RA & $1.3986\times10^{-17}$ & 0.4599 & 1865.0745 & 2000:5000\\
        $^{***}$\ch{Mg + S -> MgS + $hv$} & RA & $5.2886\times10^{-16}$ & -0.4931 & 6689.8221 & 5000:10000\\
        \ch{N$_2$H+ + MgS -> HMgS$^+$ + N$_2$} & IN & $7.30\times10^{-10}$ & -0.50 & 0.0 & 10:41000\\
        \ch{MgH$_2$ + S$^+$ -> MgSH$^+$ + H} & IN & $1.0\times10^{-09}$ & 0.0 & 0.0 & 10:41000\\
        \ch{MgH$_2$ + S$^+$ -> HMgS$^+$ + H} & IN & $1.0\times10^{-09}$ & 0.0 & 0.0 & 10:41000\\
        \ch{HMgS$^+$ + NH$_3$ -> MgS + NH$_4$$^+$} & IN & $1.0\times10^{-09}$ & 0.0 & 0.0 & 10:41000\\
        $^{**}$\ch{MgH + HS -> MgS + H$_2$} & NN & $8.31\times10^{-10}$ & -0.26 & 5.1 & 30:400
        \label{tab:reac_MGS}
\end{longtable}
\begin{tablenotes}
\footnotesize
\item Note 1. Reactions marked with * are those revised in a first stage, and whose parameters have been modified employing the NIST database.
\item Note 2. Reactions from \cite{bell25} are marked with **. Calculation of rate constant in section \ref{sec:constan_rates}.
\item Note 3. Reactions from \cite{wei25} are marked with ***.
\item Note 4. Reactions marked with **** are those revised in a second stage, for which quantum chemical calculations have been carried out (Table~\ref{tab:new_reac_MGS}).
\item Note 5. Type reactions account for: CE (Charge Exchange), CR (Cosmic-Ray Induced), DR (Dissociative Recombination), RR (Radiative Recombination), IN (ion-neutral), NN (neutral-neutral), PH (Photodissociation) and RA (Radiative Association).
\end{tablenotes}
\end{ThreePartTable}

\begin{table}[h]
    \centering
    \caption{Grain-surface chemical reactions added for MgS.}
    \begin{tabular}{l c c c c} 
        \toprule
        {Reaction} & {Type} & {$\alpha$} & {$\beta$} & {$\gamma$}\\
        \midrule
        \ch{HMgS$^+$ -> \# MgS + H} & FREEZE & 1 & 1 & 0\\
        \bottomrule
    \end{tabular}
    \label{tab:chem_mgs_grain}
\end{table}

\subsubsection{Quantum chemical computations}\label{sec:quantum_calculations}

CCSD(T)-F12/cc-pVTZ-F12 geometry optimisations and zero-point vibrational frequencies are computed with the MOLPRO2025.1 software \citep{Adler2007, Knizia2009, Peterson2008,Werner2012} for the set of reactions listed in Tables~\ref{tab:second_stage_reac_MGS} and~\ref{tab:new_reac_MGS}. Reactions in Table~\ref{tab:second_stage_reac_MGS} are already included in the chemical network by analogy with SiS (Table~\ref{tab:new_reac_MGS}), while reactions in table~\ref{tab:new_reac_MGS} correspond to newly proposed pathways included in the updated network developed herein.

\begin{ThreePartTable}
\begin{longtable}{l c c c c c c}
    \caption{PES values for the gas-phase chemical reactions for MgS. The coefficients $\alpha$, $\beta$, $\gamma$ follow the Kooij-Arrhenius equation $k(\rm T) = \alpha (\frac{T}{300K})^{\beta}\rm exp (-\gamma/T)$, where T corresponds to the temperature range included in the last column. } \\
    \toprule
    {Reaction} & {Type} & {$\alpha$} & {$\beta$} & {$\gamma$} & {Temperature range (K)} & {PES (kcal mol$^{-1}$)}\\
    \midrule
    \endfirsthead

    \multicolumn{6}{c}{\textit{Continued from previous page}} \\
    \toprule
    {Reaction} & {Type} & {$\alpha$} & {$\beta$} & {$\gamma$} & {Temperature range (K)} & {PES (kcal mol$^{-1}$)}\\
    \midrule
    \endhead

    \midrule \multicolumn{6}{r}{\textit{Continued on next page}} \\
    \endfoot

    \bottomrule
    \endlastfoot
        \ch{Mg$^+$ + OCS -> MgS$^+$ + CO} & {IN} & {0.0} & {0.0} & {0.0} & {10:41000} & 25.44\\
        \ch{Mg + HS -> MgS + H} & {NN} & {0.0} & {0.0} & {0.0} & {10:41000} & 32.34\\
        \ch{Mg + H$_2$S -> MgS + H$_2$} & {NN} & {0.0} & {0.0} & {0.0} & {10:300} & 19.47\\
        \ch{Mg + H$_2$S -> HMgS + H} & {NN} & {0.0} & {0.0} & {0.0} & {10:300} & 63.35\\
        \ch {MgS$^+$ + e$^-$ -> Mg + S} & {DR} & {2.00$\times10^{-7}$} & {-0.50} & {0.0} & {10:300} & -126.26
        \label{tab:second_stage_reac_MGS}
\end{longtable}
\begin{tablenotes}
\footnotesize
\item Note 1. $\alpha$, $\beta$ and $\gamma$ parameters for the favourable reactions are educated guesses.
\item Note 2. Type reactions account for: DR (Dissociative Recombination), RR (Radiative Recombination), IN (ion-neutral) and NN (neutral-neutral).
\end{tablenotes}
\end{ThreePartTable}

\begin{ThreePartTable}
\begin{longtable}{l c c c c c c}
    \caption{PES values for the new gas-phase chemical reactions added for MgS. The coefficients $\alpha$, $\beta$, $\gamma$ follow the Kooij-Arrhenius equation $k(\rm T) = \alpha (\frac{T}{300K})^{\beta}\rm exp (-\gamma/T)$, where T corresponds to the temperature range included in the last column.} \\
    \toprule
    {Reaction} & {Type} & {$\alpha$} & {$\beta$} & {$\gamma$} & {Temperature range (K)} & {PES (kcal mol$^{-1}$)}\\
    \midrule
    \endfirsthead

    \multicolumn{6}{c}{\textit{continued from previous page}} \\
    \toprule
    {Reaction} & {Type} & {$\alpha$} & {$\beta$} & {$\gamma$} & {Temperature range (K)} & {PES (kcal mol$^{-1}$)}\\
    \midrule
    \endhead

    \midrule \multicolumn{6}{r}{\textit{Continued on next page}} \\
    \endfoot

    \bottomrule
    \endlastfoot
        {\ch{Mg$^+$ + HS -> MgS$^+$ + H}} & {IN} & {0.0} & {0.0} & {0.0} & {10:41000} & 35.81\\
        {\ch{Mg$^+$ + HS -> MgH+ + S}} & {IN} & {0.0} & {0.0} & {0.0} & {10:41000} & 37.17\\
        {\ch{Mg$^+$ + HS -> MgH + S$^+$}} & {IN} & {0.0} & {0.0} & {0.0} & {10:41000} & 163.65\\
        {\ch{Mg + HS$^+$ -> MgH+ + S}} & {IN} & {1.0$\times10^{-9}$} & {0.0} & {0.0} & {10:41000} & -27.92\\
        {\ch{Mg + HS$^+$ -> Mg$^+$ + HS}} & {IN} & {1.0$\times10^{-9}$} & {0.0} & {0.0} & {10:41000} & -65.09\\
        {\ch{Mg + HS$^+$ -> MgS$^+$ + H}} & {IN} & {1.0$\times10^{-9}$} & {0.0} & {0.0} & {10:41000} & -29.29\\
        {\ch{MgH+ + HS -> MgS$^+$ + H$_2$}} & {IN} & {1.0$\times10^{-9}$} & {0.0} & {0.0} & {10:41000} & -21.07\\
        {\ch{Mg$^+$ + H$_2$S -> MgS$^+$ + H$_2$}} & {NN} & {0.0} & {0.0} & {0.0} & {10:300} & 22.9\\
        {\ch{MgH + S -> MgS + H}} & {NN} & {1.0$\times10^{-10}$} & {0.0} & {0.0} & {100:2500} & -20.16\\
        {\ch{MgH + S$_2$ -> MgS + SH}} & {NN} & {1.0$\times10^{-10}$} & {0.0} & {0.0} & {100:2500} & -2.13\\
        {\ch{Mg + HS$_2$ -> MgS + HS}} & {NN} & {0.0} & {0.0} & {0.0} & {10:40000} & 22.98\\
        {\ch{Mg+ + HS$_2$ -> MgS+ + HS}} & {IN} & {0.0} & {0.0} & {0.0} & {10:40000} & 26.45\\
        {\ch{Mg + S$_2$ -> MgS + S}} & {NN} & {0.0} & {0.0} & {0.0} & {10:40000} & 50.37\\
        {\ch{Mg+ + S$_2$ -> MgS+ + S}} & {IN} & {0.0} & {0.0} & {0.0} & {10:40000} & 53.85\\
        {\ch{Mg+ + CS -> MgS+ + C}} & {IN} & {0.0} & {0.0} & {0.0} & {10:40000} & 121.04\\
        {\ch{Mg+ + CS -> MgS + C+}} & {NN} & {0.0} & {0.0} & {0.0} & {10:40000} & 202.58\\
        {\ch{Mg + CS -> MgS + C}} & {NN} & {0.0} & {0.0} & {0.0} & {10:40000} & 117.57\\
        {\ch{Mg + NCS -> MgS + CN}} & {NN} & {0.0} & {0.0} & {0.0} & {10:40000} & 75.52\\
        {\ch{Mg+ + NCS -> MgS+ + CN}} & {IN} & {0.0} & {0.0} & {0.0} & {10:40000} & 78.99\\
        {\ch{Mg + HCNS -> MgS + HCN}} & {NN} & {0.0} & {0.0} & {0.0} & {10:40000} & 7.51\\
        {\ch{Mg+ + HCNS -> MgS+ + HCN}} & {IN} & {0.0} & {0.0} & {0.0} & {10:40000} & 10.98\\
        {\ch{Mg + HNCS -> MgS + HNC}} & {NN} & {1.0$\times10^{-10}$} & {0.0} & {0.0} & {10:40000} & -44.27\\
        {\ch{Mg+ + HNCS -> MgS$^+$ + HNC}} & {IN} & {1.0$\times10^{-9}$} & {0.0} & {0.0} & {10:40000} & -40.80\\
        {\ch{Mg + HCCS -> MgS + CCH}} & {NN} & {0.0} & {0.0} & {0.0} & {10:40000} & 34.70\\
        {\ch{Mg+ + HCCS -> MgS+ + CCH}} & {IN} & {0.0} & {0.0} & {0.0} & {10:40000} & 38.18\\
        {\ch{MgH + HCCS -> MgS + HCCH}} & {NN} & {1.0$\times10^{-10}$} & {0.0} & {0.0} & {10:40000} & -65.73\\
        {\ch{MgH + HCCS -> MgS + CCH2}} & {NN} & {1.0$\times10^{-10}$} & {0.0} & {0.0} & {10:40000} & -22.78\\
        {\ch{MgH+ + HCCS -> MgS$^+$ + HCCH}} & {IN} & {1.00$\times10^{-9}$} & {0.0} & {0.0} & {10:40000} & -46.93\\
        {\ch{Mg + H2CCS -> MgS + H2CC}} & {NN} & {0.0} & {0.0} & {0.0} & {10:40000} & 64.52\\
        {\ch{Mg+ + H2CCS -> MgS+ + H2CC}} & {IN} & {0.0} & {0.0} & {0.0} & {10:40000} & 67.99\\
        {\ch{Mg + HOCS+ -> MgS+ + HOC}} & {NN} & {0.0} & {0.0} & {0.0} & {10:40000} & 56.51\\
        {\ch{Mg + HOCS+ -> MgS + HOC+}} & {IN} & {0.0} & {0.0} & {0.0} & {10:40000} & 62.76\\
        {\ch{Mg + NS -> MgS + N}} & {NN} & {0.0} & {0.0} & {0.0} & {10:40000} & 59.84\\
        {\ch{Mg+ + NS -> MgS+ + N}} & {IN} & {0.0} & {0.0} & {0.0} & {10:40000} & 63.31\\
        {\ch{Mg + SiS -> MgS + Si}} & {NN} & {0.0} & {0.0} & {0.0} & {10:40000} & 93.71\\
        {\ch{Mg+ + SiS -> MgS+ + Si}} & {IN} & {0.0} & {0.0} & {0.0} & {10:40000} & 97.19\\
        {\ch{S + H$_3$+ -> SH$^+$ + H$_2$}} & {IN} & {1.0$\times10^{-9}$} & {0.0} & {0.0} & {10:40000} & -57.85\\
        {\ch{S + HCO+ -> SH$^+$ + CO}} & {IN} & {1.0$\times10^{-9}$} & {0.0} & {0.0} & {10:40000} & -17.38\\
        {\ch{S + N$_2$H+ -> SH$^+$ + N$_2$}} & {IN} & {1.0$\times10^{-9}$} & {0.0} & {0.0} & {10:40000} & -40.76\\
        {\ch{Mg + H$_3$+ -> MgH$^+$ + H$_2$}} & {IN} & {1.0$\times10^{-9}$} & {0.0} & {0.0} & {10:40000} & -94.96\\
        {\ch{Mg + HCO+ -> MgH$^+$ + CO}} & {IN} & {1.0$\times10^{-9}$} & {0.0} & {0.0} & {10:40000} & -54.49\\
        {\ch{Mg + N$_2$H+ -> MgH$^+$ + N$_2$}} & {IN} & {1.0$\times10^{-9}$} & {0.0} & {0.0} & {10:40000} & -77.87\\
        {\ch{Mg + H$_3$O+ -> MgH$^+$ + H$_2$O}} & {IN} & {1.0$\times10^{-9}$} & {0.0} & {0.0} & {10:40000} & -30.69\\
        {\ch {MgS$^+$ + e$^-$ -> MgS + $hv$}} & {RR} & {2.0$\times10^{-10}$} & {-0.50} & {0.0} & {10:300} & -117.14\\
        {\ch {Mg$^+$ + S -> MgS + $hv$}} & {RA} & {1.0$\times10^{-18}$} & {0.0} & {0.0} & {10:40000} & -47.41
        \label{tab:new_reac_MGS}
\end{longtable}
\begin{tablenotes}
\footnotesize
\item Note 1. $\alpha$, $\beta$ and $\gamma$ parameters for the favourable reactions are educated guesses.
\item Note 2. Type reactions account for: DR (Dissociative Recombination), RR (Radiative Recombination), IN (ion-neutral), NN (neutral-neutral) and RA (Radiative association).
\end{tablenotes}
\end{ThreePartTable}

\subsubsection{Calculation of the rate constant for reaction \ch{MgH + HS -> MgS + H$_2$}}\label{sec:constan_rates}
Global rate constant for the reaction between MgH and HS leading to MgS and H$_2$ was calculated between 30 and 400 K and at 1$\times$10$^{-5}$ atm based on the PES previously computed by \cite{bell25}. For the elementary step ruled by the transition state of the dissociation of HMgSH the unimolecular rate coefficient was calculated using Rice-Ramsperger-Kassel-Marcus (RRKM) theory within the rigid-rotor harmonic-oscillator (RRHO) approximation (Weston 1972). The barrierless association reaction between MgH and HS was obtained employing the phase space theory \citep{Pechukas1965, Chesnavich1986}, where the attractive potential between the two fragments at large distances is described by a $V(R)=-Cn/R^n$ functional form. The $Cn$ constant is derived from a fit of the B3LYP/aug-cc-pVTZ energies along the Mg-S distance. We analysed the asymptotic behaviour of the attractive interaction by testing different inverse-power laws. Effective coefficients were defined as $Cn(R)=-V(R)*R^n$ for n=1-6. A clear plateau region was obtained exclusively for n=3. This behaviour indicates that the interaction along the relaxed minimum-energy path is dominated by an effective $R^3$ term. On this basis, the $V(R)=-C/R^3$ functional was adopted to model the barrierless association process. Temperature- and pressure-dependent phenomenological rate coefficients were subsequently evaluated within a master equation formalism based on transition state theory (AITSTME). The master equation was solved using the MESS software package \citep{Georgievskii2013}.

To describe the temperature dependence of the global rate constant, the kinetic results were fitted to the Arrhenius-modified expression (Arrhenius–Kooij equation), were derived alpha, beta and gamma parameters are 8.31$\times$10$^{-10}$ cm$^3$/molec*s, -0.26 and 5.10 K, respectively.

\newpage
\section{sulphur constrains}
\subsection{Final temperature = 10 K}\label{sec:s_analy}
\begin{figure}[h]
    \centering
    \includegraphics[width=\linewidth]{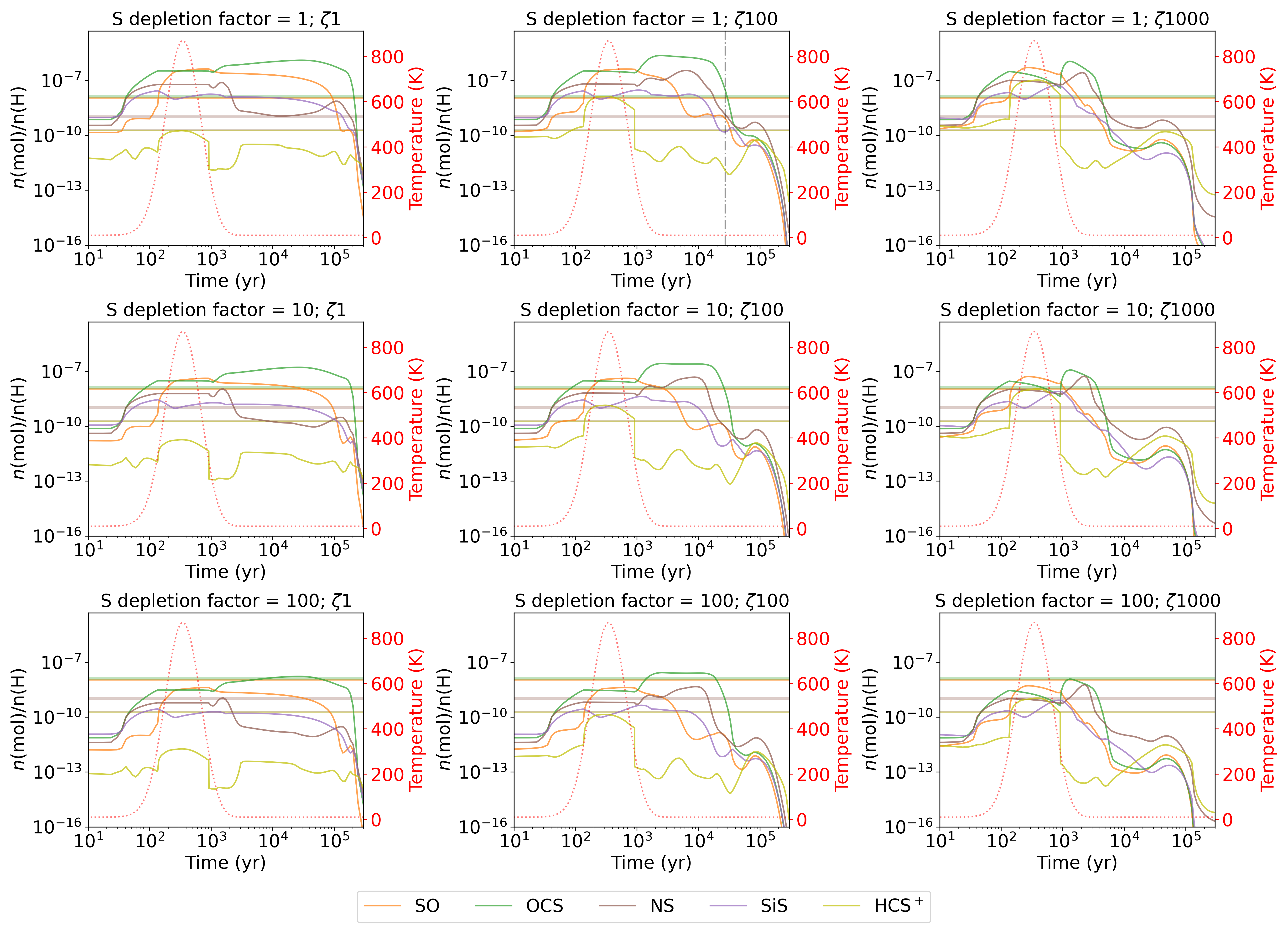}
    \caption{Chemical model results (Phase 2): evolution of the fractional abundances of SO, OCS, NS, SiS and HCS$^+$ in the gas phase as a function of time. We consider for this phase a C-type shock with a shock speed of $c_s$=20 km s$^{-1}$ and an initial gas density of $n(\rm H)$=2$\times$10$^4$ cm$^{-3}$. 
    We explore three different values for the cosmic-ray ionisation rate: $\zeta$1 is the standard Galactic value of 1.3$\times$10$^{-17}$ (First column), and two enhanced values of $\zeta$ that are $\zeta$100 which is 100 times higher than the standard one (middle column) and $\zeta$1000 that is 1000 times higher than the standard one (last column). In addition, we run the model for three depletion factors that are no depletion (i.e, S depletion factor = 1) in the top row; a depletion factor of 10 shown in the middle row; and a depletion factor of 100 in the bottom row.
    The grey dashdotted line corresponds to 2.7$\times$10$^4$ years as inferred using the MAE method and indicates the timescale where to compare the observed abundances. The red dashed line indicates the evolution of the temperature of the neutral fluid within the shock. Finally, the shaded horizontal lines correspond to the observed abundances towards G+0.693, adopting $n$(H)=2$\times$$n$(H$_2$), taking into account their error, and are colour-coded according to the legend.}
    \label{fig:S_constrain_todos}
\end{figure}
The MAE analysis serves as a minimisation metric, where the model yielding the smallest MAE value is, by definition, the best‑fitting model. For the sulphur analysis, the minimum MAE obtained is 0.774$^{+0.014} _{-0.015}$, corresponding to the parameter combination D$_1$ and $\zeta_2$, as reported in Table~\ref{tab:S_MAE}. This value remains the global minimum within its associated uncertainties and does not overlap with the next‑lowest MAE value, indicating that it represents a robust and well‑constrained solution.
When compared with the next best model (MAE=0.829), the minimum MAE is 6\% lower, further supporting the reliability of this solution.

\begin{table}[h]
    \centering
    \caption{MAE values for the grid of models. $\zeta$ is the cosmic-ray ionisation rate. D is the depletion factor of S.}
    \begin{tabular}{c c c c}
        \toprule
        {$\zeta$/D} & {D$_1$=0} & {D$_2$=10} & {D$_3$=100}\\
        \midrule
        
        {$\zeta$$_1$=1} & 0.829$^{+0.013} _{-0.014}$ & 0.930$\pm 0.015$ & 1.096$\pm 0.014$\\
        
        \hline
        {$\zeta$$_2$=100} & 0.774$^{+0.014} _{-0.015}$ & 1.266$\pm 0.014$ & 1.614$\pm 0.015$\\
        
        \hline
        {$\zeta$$_3$=1000} & 1.271$^{+0.014} _{-0.015}$ & 2.076$\pm 0.015$ & 3.052$\pm 0.015$\\
        \bottomrule
    \end{tabular}
    \label{tab:S_MAE}
\end{table}

\section{Magnesium constrains}
\subsection{Final temperature = 10 K}\label{sec:mg_analy}
\begin{figure}[h]
    \centering
    \includegraphics[width=\linewidth]{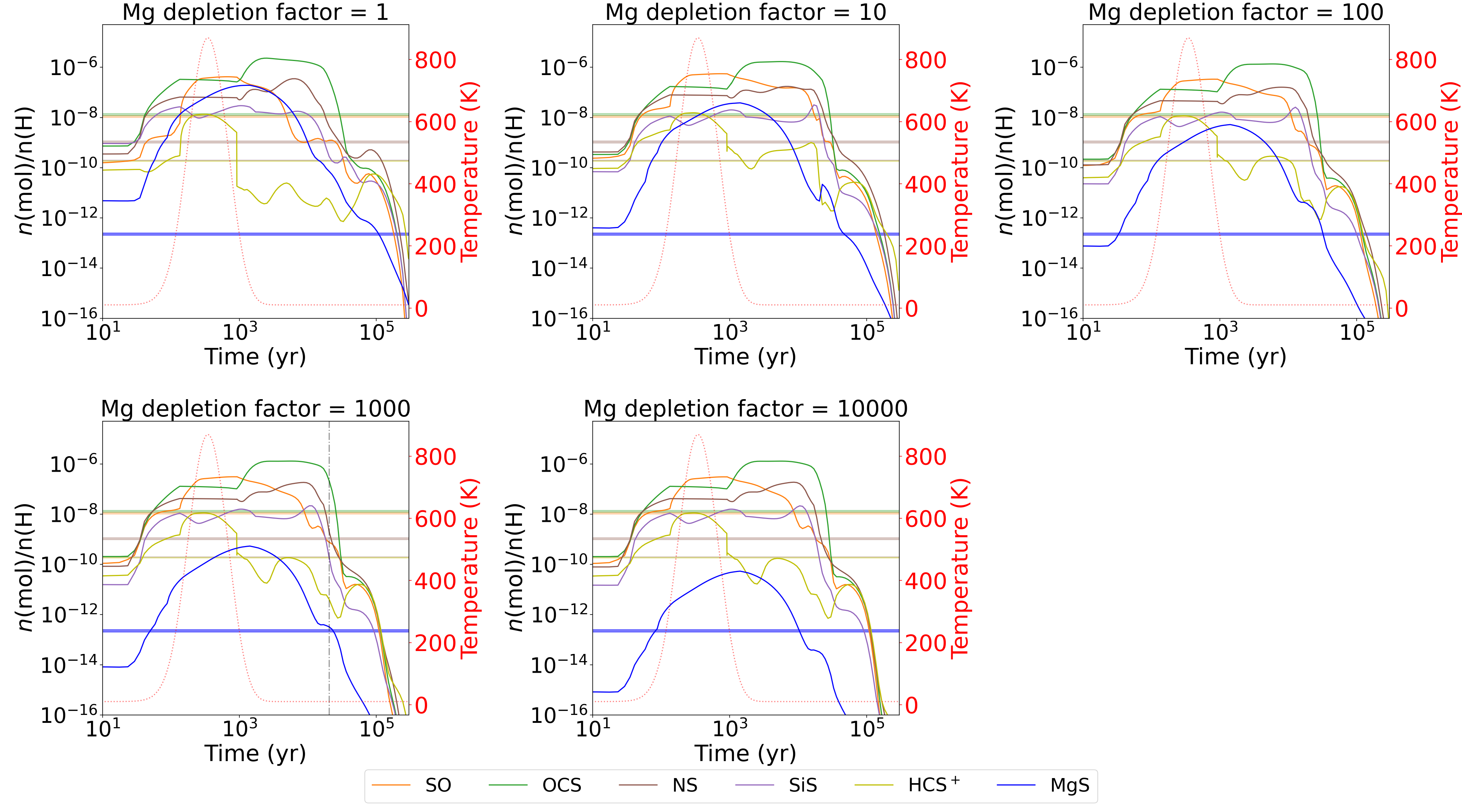}
    \caption{Chemical model results (Phase 2): evolution of the fractional abundances of SO, OCS, NS, SiS, HCS$^+$ and MgS in the gas phase as a function of time. We also explore five different depletion factors for the initial abundances of Mg: the initial elemental abundance reported for this specie in \cite{Asplund2009} 3.98$\times$10$^{-5}$ (first row, left panel), and four enhanced values that are 10 times (first row, middle panel), 100 times (first row, right panel), 1000 times (second row, left panel) and 10000 times (second row, middle panel) lower than the standard one. Finally, the shaded horizontal line corresponds to the observed abundances towards G+0.693, adopting $n$(H)=2$\times$$n$(H$_2$), taking into account their error, and are colour-coded according to the legend at the bottom of the figure. The grey dashdotted line corresponds to 2.1$\times$10$^4$ years as inferred using the MAE method and indicates the timescale where to compare the observed abundances. The red dashed line indicates the evolution of the temperature of the neutral fluid within the shock.}
    \label{fig:Mg_constrain}
\end{figure}
The minimum MAE value is 0.750$\pm 0.015$, obtained for D$_4$, corresponding to a depletion factor of 1000 (Table~\ref{tab:MG_MAE}). Its uncertainty interval does not intersect with that of the next‑lowest MAE, indicating that it is a statistically distinct and well‑defined minimum.
Quantitatively, the minimum MAE is 14\% lower than the next best value, further reinforcing the reliability of this result.

\begin{table}[h]
    \centering
    \caption{MAE values for the grid of models. $\zeta$ is the cosmic-ray ionisation rate. D is the depletion factor for Mg.}
    \begin{tabular}{c c c c c c}
        \toprule
        {$\zeta$/D} & {D$_1$=0} & {D$_2$=10} & {D$_3$=100} & {D$_4$=1000} & {D$_5$=10000}\\
        \midrule
        {$\zeta$=100} & 0.960$^{+0.015} _{-0.014}$ & 0.865$\pm 0.015$ & 0.882$^{+0.014} _{-0.015}$ & 0.750$\pm 0.015$ & 0.864$\pm 0.015$\\
        \bottomrule
    \end{tabular}
    \label{tab:MG_MAE}
\end{table}

\subsection{Final temperature = 100 K}\label{sec:mg_analy_tf100}
\begin{table}[h]
    \centering
    \caption{Chemical reactions and their percentage of importance, divided into formation and destruction of MgS.}
    \begin{tabular}{l c}
        \toprule
        {Reaction} & {Percentage \%} \\
        \midrule
        \multicolumn{2}{c}{{Formation Reactions}} \\
        \midrule
        \ch{MgH + S -> MgS + H} & 89.6 \\
        \ch{MgSH$_2$ + C -> MgS + CH$_2$} & 5.5 \\
        \ch{MgH + HS -> MgS + H$_2$} & 2.1 \\
        \ch{HMgS$^+$ + NH$_3$ -> MgS + NH$_4 ^+$} & 0.96 \\
        \ch{MgH + HCCS -> MgS + CCH$_2$} & 0.65 \\
        \ch{MgH + HCCS -> MgS + HCCH} & 0.65 \\
        \ch{HCN + HMgS$^+$ -> MgS + HCNH$^+$} & 0.27 \\
        \ch{MgH + S$_2$ -> MgS + SH} & 0.16 \\
        $^a$\ch{\#MgS + DEUVCR -> MgS} & 0.04\\
        $^b$\ch{\@MgS + THERM -> MgS} & 0.03\\
        \midrule
        \multicolumn{2}{c}{{Destruction Reactions}} \\
        \midrule
        \ch{H$_3$$^+$ + MgS -> HMgS+ + H$_2$} & 45.4\\
        \ch{S$^+$ + MgS -> MgS$^+$ + S} & 12.9 \\
        \ch{H$^+$ + MgS -> MgS$^+$ + H} & 9.5 \\
        \ch{HCO+ + MgS -> HMgS+ + CO} & 9.4 \\
        \ch{He$^+$ + MgS -> S$^+$ + Mg + He} & 6.5 \\
        \ch{He$^+$ + MgS -> S + Mg$^+$ + He} & 6.5 \\
        \ch{MgS + CRPHOT -> S + Mg} & 4.1 \\
        \ch{C$^+$ + MgS -> MgS$^+$ + S} & 2.0 \\
        \ch{C$^+$ + MgS -> MgC$^+$ + C} & 2.0 \\
        \ch{MgS + FREEZE -> \#MgS} & 1.3 \\
        \bottomrule
    \end{tabular}
    \label{tab:MgS_Tf_100}
\begin{tablenotes}
\footnotesize
\item $^a$ \#MgS denotes surface ice MgS.
\item $^b$ \@MgS denotes bulk ice MgS.
\end{tablenotes}
\end{table}

\end{appendix}

\end{document}